\pdfoutput=1 
\documentclass[pre,footinbib,twocolumn,final,superscriptaddress]{revtex4}

\usepackage{amssymb,amsbsy,amsfonts,mathrsfs,amsmath}

\usepackage{bbm}

\usepackage[T1]{fontenc} 
\usepackage[english,francais]{babel}

\usepackage{graphicx,epsfig,color}

\usepackage{url,hyperref}  
\hypersetup{
      colorlinks=true,
      linkcolor=blue,
      filecolor=blue,
      citecolor = red,
      urlcolor=blue,
    }

\usepackage{enumitem}

\definecolor{M_Beige}         {rgb}{0.96 , 0.96 , 0.86}

\definecolor{M_Brown}         {rgb}{0.65 , 0.16 , 0.16}

\definecolor{M_Gold}          {rgb}{1.00 , 0.84 , 0.00}

\definecolor{M_LemonChiffon}  {rgb}{1.00 , 0.98 , 0.80}

\definecolor{M_Orange}        {rgb}{1.00 , 0.60 , 0.00}

\definecolor{M_Pink}          {rgb}{0.80 , 0.55 , 0.60}

\definecolor{M_Violet}          {rgb}{0.83 , 0.21 , 0.93}

\definecolor{M_Green}          {rgb}{0.2 , 0.6 , 0.2}

\definecolor{M_Gray}          {rgb}{0.5 , 0.5 , 0.5}

\definecolor{M_BluPal}          {rgb}{0.7 , 0.7 , 0.9}

\def\XXint#1#2#3{{\setbox0=\hbox{$#1{#2#3}{\int}$}
\vcenter{\hbox{$#2#3$}}\kern-.5\wd0}}

\renewcommand{\leq}{\leqslant}
\renewcommand{\geq}{\geqslant}

\newcommand{\EXP}[1]{\mathrm{e}^{#1}}         

\def\eqdef{\stackrel{\mbox{\tiny def}}{=}}     
\newcommand{\ket}[1]{|\kern.3ex#1\kern.3ex\rangle}
\newcommand{\bra}[1]{\langle\kern.3ex #1 \kern.3ex|}
\newcommand{\mean}[1]{\left\langle #1 \right\rangle} 
\newcommand{\smean}[1]{\langle #1 \rangle} 

\newcommand{\symmetriz}[1]{\left\lfloor #1 \right\rfloor}
\newcommand{\ssymmetriz}[1]{\lfloor #1\rfloor}

\renewcommand{\min}[2]{\mathop{\mathrm{min}}\nolimits\left( #1 , #2\right)}

\newcommand{\heaviside}{\mathop{\theta_\mathrm{H}}\nolimits}  

\def\I{{\rm i}}                  
\def\D{{\rm d}}                  

\newcommand{\deriv}[2]{\frac{\mathrm{d}#1}{\mathrm{d}#2}}
\newcommand{\derivp}[2]{\frac{\partial #1}{\partial #2}}

\newcommand\antiddots{\mathinner{\mkern2mu\raise1pt\hbox{.}\mkern2mu
    \newline \raise4pt\hbox{.}\mkern2mu\raise7pt\hbox{.}\mkern1mu}}

\def\IDoS{\mathcal{N}}

\def\K{K}
\def\mb{\overline{m}}
\def\Kb{\tilde{K}}

\def\rh{t}

\begin{document}

\renewcommand{\labelitemi}{$\bullet$}
\renewcommand{\labelitemii}{$\star$}

\selectlanguage{english}

\title{Disordered harmonic chains with random masses and springs: \\ a combinatorial approach}

\author{Maximilien Bernard}
\affiliation{LPTMS, Universit\'e Paris-Saclay, CNRS, 91405 Orsay, France}
\affiliation{Laboratoire de Physique de l’Ecole Normale Supérieure,
CNRS, ENS and Universit\'e PSL, Sorbonne Universit\'e,
Université Paris Cit\'e, 24 rue Lhomond, 75005 Paris, France}
\author{Christophe Texier}
\affiliation{LPTMS, Universit\'e Paris-Saclay, CNRS, 91405 Orsay, France}

\date{January 21, 2026}


%
%




\begin{abstract}
We study harmonic chains with i.i.d. random spring constants $K_n$ and i.i.d. random masses $m_n$.
We introduce a new combinatorial approach which allows to derive a compact and general approximate formula for the complex Lyapunov exponent, in terms of the solutions of two transcendental equations involving the distributions of the spring constants and the masses.
Our result makes easy the asymptotic analysis of the low frequency properties of the eigenmodes (spectral density and localization) for arbitrary disorder distribution, as well as their high frequency properties.
We apply the method to the case of power-law distributions $p(K)=\mu\,K^{-1+\mu}$ with $0<K<1$ and $q(m)=\nu\,m^{-1-\nu}$ with $m>1$ (with $\mu,\:\nu>0$).
At low frequency, the spectral density presents the power law $\varrho(\omega\to0)\sim\omega^{2\eta-1}$, where the exponent $\eta$ exhibits first order phase transitions on the line $\mu=1$ and on the line $\nu=1$.
The exponent of the non disordered chain ($\eta=1/2$) is recovered when $\langle K_n^{-1}\rangle$ and $\langle m_n\rangle$ are both finite, i.e. $\mu>1$ and $\nu>1$.
The Lyapunov exponent (inverse localization length) shows also a power-law behaviour $\gamma(\omega^2\to0)\sim\omega^{2\zeta}$, where the exponent $\zeta$ exhibits several phase transitions~:  
the exponent is $\zeta=\eta$ for $\mu<1$ or $\nu<1$ ($\langle K_n^{-1}\rangle$ or $\langle m_n\rangle$ infinite) and $\zeta=1$ when $\mu>2$ and $\nu>2$ ($\langle K_n^{-2}\rangle$ and $\langle m_n^2\rangle$ both finite). In the intermediate region it is given by $\zeta=\mathrm{min}(\mu,\nu)/2$. 
On the transition lines, $\varrho(\omega)$ and $\gamma(\omega^2)$ receive logarithmic corrections.

Finally, we illustrate the versatility of our combinatorial approach by considering the case of the Anderson model with random couplings (a model known to be mapped onto the random spring chain for ``Dyson type I'' disorder). 
\end{abstract}

\date{\today}

\maketitle


\section{Introduction}
\label{sec:Intro}

We study here a one-dimensional chain of masses coupled by springs.
The displacement $X_n(t)$ of the $n$-th mass $m_n$ obeys the Newton equation for harmonic forces~; we denote $\K_n$ the spring constant coupling masses $n$ and $n+1$.
The fundamental question is to understand the properties of the eigenmodes obtained by solving the equations of motion for $X_n(t)=x_n\,\EXP{-\I\omega t}$, 
\begin{equation}
  \label{eq:WaveEquation}
 -m_n\omega^2 x_n 
 =  \K_{n-1}\, (x_{n-1} - x_n) + \K_{n}\, (x_{n+1} - x_n) 
 \:.
\end{equation}
We consider the situation where the chain is free at one side, $\K_0=0$, and pinned at the other side, $\K_N\neq0$ and $x_{N+1}=0$. This choice of boundary conditions slightly simplifies the analysis.
The spectral problem has then $N$ solutions $\omega_\alpha^2$ for $\alpha=1,\cdots,N$.
In the pure case, $m_n=m$ and $\K_n=\K$ $\forall n$, the vibrational modes are extended and the mode density 
$\rho(\lambda)\eqdef\lim_{N\to\infty}(1/N)\sum_{\alpha=1}^N\delta(\lambda -\omega_\alpha^2)$ is given by 
$\rho(\lambda)=(1/\pi)/\sqrt{\lambda(4\omega_0^2-\lambda)}$ for $\lambda\in[0,4\omega_0^2]$, where $\omega_0^2=\K/m$~\cite{footnote1}.  
In the presence of a quenched disorder, where $m_n$ and/or $\K_n$ depend randomly on $n$, two important questions are then~:
(Q1) how the eigenvalues are distributed, i.e. how $\rho(\lambda)$ deviates from the one of the pure case~?  
(Q2) What are the localization properties of the eigenmodes~?

The random spring chain is the first model of wave localization studied in the seminal papers of Dyson \cite{Dys53} and Schmidt \cite{Sch57} (see also the reviews \cite{LieMat66,Ish73}).
The two questions (Q1) and (Q2) have been addressed analytically or numerically by many authors for random masses \cite{Sch57,Aga64,MatIsh70,Nie82,Nie84,NieLuc85,NieLucCanVanVen86,NieLuc87,NieLuc87a,LucNie88}
\cite{footnote2} 
or random spring constants \cite{Dea64}~;
the random spring constant problem was studied in other contexts involving the same spectral problem~: 1D master equation with random hopping rates \cite{BerSchWys80,AleBerSchOrb81,SteKar82}, random LC line \cite{AkkMay84}, random Laplacian \cite{GarItzDer84} and recently fluctuating interfaces \cite{BerLeDRosTex24}.
To the best of our knowledge, the case where both the masses and the spring constants are random has not been addressed, with the notable exception of Nieuwenhuizen's paper \cite{Nie84} where a specific solvable case was identified~\cite{footnote3}.

The purpose of the present article is to address the spectral and localization properties of random spring chains in the case where both masses and spring constants are independent and identically distributed (i.i.d.) with power-law distributions~:
$p(\K)=\mu\,\K^{-1+\mu}$ for $\K\in[0,1]$ and 
$q(m)=\nu\,m^{-1-\nu}$ for $m\in[1,\infty[$, respectively, with $\mu>0$ and $\nu>0$.
We distinguish (i) the case $\mu>2$ when $\mean{\K_n^{-2}}$ is finite, (ii) the case $1<\mu\leq2$ when $\mean{\K_n^{-2}}=\infty$ and $\mean{\K_n^{-1}}$ is finite, and (iii) the case $0<\mu\leq1$ when $\mean{\K_n^{-1}}=\infty$ (and the same for the mass). Here $\mean{\cdots}$ denotes disorder averaging.
We show that, for disorder uncorrelated in space, this covers all possible cases of low frequency behaviours for the spectral density and the localization length for random spring chains.


\section{Main results and outline}
\label{sec:Res}

By using a new combinatorial method, we obtain the solution of Eq.~\eqref{eq:WaveEquation}, denoted $x_{n+1}(\omega^2)$, for initial value $x_1=1$.
This solution is also the spectral determinant of the $N\times N$ tridiagonal matrix $\Lambda$ defining the spectral problem~\eqref{eq:WaveEquation}, rewritten as $\sum_m\Lambda_{nm} x_m=\omega^2x_n$~:
setting $\lambda=\omega^2$
we have
\begin{equation}
  \label{eq:MainRes0}
  \hspace{-0.25cm}
  x_{N+1}(\lambda) 
  =  \det\left( \Lambda - \lambda\,\mathbf{1}_N \right)\,\prod_{j=1}^N\frac{m_j}{\K_j}
  = 
  \sum_{k=0}^N a_{N+1}^{(k)}\,(-\lambda)^k
\end{equation}
Our first result is an exact representation of the coefficients in terms of multiple sums
\begin{equation}
  \label{eq:MainResIntro2}
  a_{n+1}^{(k)} 
  = 
  \hspace{-0.5cm}
  \sum_{1\leq j_1 \leq i_1< j_2 \leq i_2<\cdots < j_k\leq i_k\leq n}
  \   \prod_{m=1}^k \frac{m_{j_m}}{K_{i_m}}
\end{equation}
Assuming some symmetrization with respect to the disorder parameters allows some 
precise
asymptotic analysis of these coefficients, which leads to a new compact and general formula for the complex Lyapunov exponent
\begin{align}
  \label{eq:MainResultForOmegaIntro}
  \Omega(\lambda) \simeq - \underset{\theta}{\mathrm{min}}
  \int_0^\theta\D t\, \ln\left[
    \frac{4}{(-\lambda)}\,\frac{1-t}{1+t}\,\varphi_*^{(1/\K)}(t)\varphi_*^{(m)}(t)
  \right]
  \:.
\end{align}
where the ``fugacity'' $\varphi_*^{(1/\K)}(\theta)$ is the solution of the transcendental equation
\begin{equation}
  \label{eq:Transcendental-K}
  \varphi_*\,\mean{\frac{1/\K_i}{1+\varphi_*/\K_i}} = \theta
  \:,
\end{equation}
and the ``fugacity'' $\varphi_*^{(m)}(\theta)$ is solution of 
\begin{equation}
  \label{eq:Transcendental-m}
  \varphi_*\,\mean{\frac{m_j}{1+\varphi_*\,m_j}} = \theta
  \:.
\end{equation}
Although Eq.~\eqref{eq:MainResultForOmegaIntro} is not exact as it relies on a symmetrization of the coefficients \eqref{eq:MainResIntro2}, its interest is to provide straightforwardly the limiting behaviours of $\Omega(\lambda)$ for $\lambda<0$, for arbitrary distributions of spring constants and masses.
Some analytic continuation provides the spectral density and the Lyapunov exponent.
For power-law distributions of spring constants and masses, the spectral density presents the power-law behaviour
$\rho(\lambda)\sim\lambda^{\eta-1}$ for $\lambda\to0^+$ 
with exponent $\eta$ given in Table~\ref{tab:Eta}.

\begin{table}[!h]
\centering
\begin{tabular}{cc|cc|}
     $\eta$: &                   &  $\mu<1 $               & $\mu>1 $
           \\
           &                     & $\mean{\K_n^{-1}}=\infty$    & $\mean{\K_n^{-1}}<\infty$
       \\ [0.125cm]
       \hline
   $\nu<1$ & $\mean{m_n}=\infty$ &  $\big(\frac{1}{\mu} +\frac{1}{\nu} \big)^{-1}$ & $\frac{\nu}{1+\nu}$  
   \\[0.125cm]
   $\nu>1$ & $\mean{m_n}<\infty$ &  $\frac{\mu}{1+\mu}$ & $1/2$
   \\
   \hline
\end{tabular}
\caption{\it Exponent $\eta$ controlling the spectral density at low frequency.}
\label{tab:Eta}
\end{table}


The Lyapunov exponent, measuring the localization of the modes, presents a power-law behaviour 
$\gamma(\lambda)\sim(-\lambda)^{\eta}$ for $\lambda\to0^-$ (evanescent modes).
For  $\lambda\to0^+$ (propagating modes), the discussion is more complex~:
we 
write
 $\gamma(\lambda)\sim \lambda^{\zeta}$ for $\lambda\to0^+$.
The exponents coincide, $\zeta=\eta$, only when $\mu<1$ or $\nu<1$.
For $\mu>2$ and $\nu>2$ ($\mean{\K_n^{-2}}<\infty$ and $\mean{m_n^2}<\infty$), we recover $\zeta=1$ (standard perturbative result).
In the intermediate region we get $\zeta=(1/2)\min{\mu}{\nu}$. These results are summarized in Table~\ref{tab:Zeta}.

The exponents $\eta$ and $\zeta$ are continuous on the transition lines, however additional logarithmic corrections appear there, given in Section~\ref{sec:LowFreq}.

\begin{table}[!h]
\centering
\begin{tabular}{l|ccc|}
   $\zeta$\,:    
     & $0<\mu<1$               
     & $1<\mu<2$              
     & $\mu>2$
   \\ 
   \hline
   $0<\nu<1$   
     & $\big(\frac{1}{\mu} +\frac{1}{\nu} \big)^{-1} $
     & $\big( 1 +\frac{1}{\nu} \big)^{-1}$ 
     & $\big( 1 +\frac{1}{\nu} \big)^{-1} $ 
   \\[0.125cm]
   $1<\nu<2$   
     & $\big(\frac{1}{\mu} +1 \big)^{-1}$  
     & $\frac12\min{\mu}{\nu} $                         
     & $\frac{\nu}{2}$
   \\[0.125cm]
   $\nu>2$     
     & $\big(\frac{1}{\mu} +1 \big)^{-1} $ 
     & $\frac{\mu}{2}$             
     & $1$
   \\
   \hline
\end{tabular}
\caption{\it Exponent $\zeta$ controlling the low frequency Lyapunov exponent (inverse localization length).}
\label{tab:Zeta}
\end{table}

The plan of the paper is as follows~:
in Section~\ref{sec:ComplexLyap}, we define the complex Lyapunov exponent of the 
disordered one-dimensional spring chain.
Section~\ref{sec:Duality} discusses the symmetry of random spring chains, which will be used at several places.
In Section~\ref{sec:Combinatorics}, we derive the exact solution of the initial value problem, i.e. Eq.~\eqref{eq:MainResIntro2}.
Its asymptotic analysis for large $n$ leads to the compact approximate expression \eqref{eq:MainResultForOmegaIntro}. 
Based on this result, we show how asymptotic analysis allows to deduce straightforwardly the limiting behaviours of the complex Lyapunov exponent.
In Section~\ref{sec:LowFreq}, we apply the method introduced 
to the model with power-law disorder and classify the low energy spectral and localization properties of the chain for different types of disorder.
In Section~\ref{sec:DysonTypeI}, we show that the method is not restricted to the spring chain model by applying the same idea to the Anderson model with random hoppings.
After some concluding remarks (Section~\ref{sec:Conclu}), the paper ends with several appendices providing some technical details on the calculations or describing other versions of the spring chain model (the discrete model for different boundary conditions 
and
 a continuous version of the model).


\section{The complex Lyapunov exponent}
\label{sec:ComplexLyap}

One dimensional wave equations with disorder can be studied thanks to powerful analytical methods, known as the Dyson-Schmidt method \cite{Dys53,Sch57,Luc92} in discrete setting, or the phase formalism \cite{AntPasSly81,LifGrePas88} for continuous models.
The method relies on the study of certain stochastic processes thanks to standard techniques (stochastic differential equations, Fokker-Planck equation,...).
The method can be equivalently formulated in terms of transfer matrices (Section~\ref{sec:Duality}), leading to consider random matrix products which can be studied within the Furstenberg theory (the integral Furstenberg equation corresponds to the Dyson-Schmidt equation).
See \cite{ComTexTou13} for a recent review of the method and \cite{ComTexTou22} for a review on exact results.
Although our starting point is the same, based on Lyapunov exponent analysis, we will \textit{not} use here the standard statistical analysis allowing to obtain the Lyapunov exponent from the expectation of a certain stochastic process.
Instead, we will deduce the Lyapunov exponent from the typical behaviour of the exact solution of the disordered wave equation, obtained by combinatorial considerations in Section~\ref{sec:Combinatorics}.

The vibrational modes of the chain of $N$ masses are given by solving the spectral problem 
\begin{equation}
  \label{eq:WaveEquation2}
 -m_n\lambda\, x_n =  \K_{n-1}\, x_{n-1} - (\K_{n-1}+\K_{n})\,x_n + \K_{n}\, x_{n+1}
\end{equation}
where $\lambda=\omega^2$ and
for boundary conditions 
\begin{equation}
  x_1=1
  \hspace{0.5cm}\mbox{and}\hspace{0.5cm}
  x_{N+1}=0
  \:.
\end{equation}
In practice, these boundary conditions require $\K_0=0$ and $\K_N\neq0$ and 
correspond to a line with free left end and pinned at the right end (the specific choice of boundary conditions is not important in the thermodynamic limit $N\to\infty$, however this choice will slightly simplify the analysis). 
We now recall the standard method allowing to obtain the density of vibrational modes and the Lyapunov exponent characterizing their localization, which relies on the node counting method and analytic continuation (although it was already used in the pioneer work of Dyson \cite{Dys53}, the use of analytic continuation was more clearly promoted by Nieuwenhuizen \cite{Nie82} and originates in the Thouless relation~\cite{Tho72}). 
The starting point is to replace the spectral problem by the inital value problem and study the solution of the recurrence
\begin{equation}
  \label{eq:TheRecurrence}
  x_{n+1}(\lambda) 
  = \left( 1+\frac{\K_{n-1}-m_n\lambda}{\K_{n}}  \right) x_n(\lambda) - \frac{\K_{n-1} }{\K_{n}} x_{n-1}(\lambda)
\end{equation}
Starting from $x_1=1$, we find  
$x_2(\lambda) = 1 -\lambda\,m_1/\K_1$,  
$x_3(\lambda) = 1 -\lambda\,\big(m_1/\K_1+m_1/\K_2+m_2/\K_2\big)+\lambda^2\,m_1m_2/(\K_1\K_2)$, 
etc. 
Inspection of the recurrence shows that $x_{n+1}(\lambda)$ is a polynomial of degree $n$ in $\lambda$. 
The quantization equation, which determines the spectrum of the chain with $N$ masses, is $x_{N+1}(\lambda)=0$. 
Hence we can represent the solution in terms of the $N$ eigenvalues $\omega_1^2,\cdots,\omega_N^2$ as
\begin{equation}
  \label{eq:SolutionIVP}
  x_{N+1}(\lambda) = \prod_{j=1}^N\frac{m_j}{\K_j} \, \prod_{\alpha=1}^N (\omega_\alpha^2-\lambda)
  \:.
\end{equation}
Let us rewrite Eq.~\eqref{eq:WaveEquation2} as $\sum_m\Lambda_{nm} x_m=\lambda\, x_n$, in terms of the $N\times N$ tridiagonal matrix $\Lambda$ defined as 
$\Lambda_{n,n-1} = -{\K_{n-1}}/{m_n}$,
$\Lambda_{n,n} = {(\K_{n-1}+\K_{n})}/{m_n}$ and $\Lambda_{n,n+1} = -{\K_{n}}/{m_n}$.
The solution $x_{N+1}(\lambda) $ of the initial value problem is also the ``\textit{spectral determinant}'' 
\begin{equation}
  \label{eq:SpectralDet-Neumann/Dirichlet}
  x_{N+1}(\lambda)  =  \prod_{j=1}^N\frac{m_j}{\K_j}
  \:\det\left( \Lambda - \lambda\,\mathbf{1}_N \right)
\end{equation}
where $\mathbf{1}_N$ is the $N\times N$ identity matrix 
%
(for further information on spectral determinants, see Appendix A of Ref.~\cite{FyoLeDRosTex18} and references therein, or the review \cite{ComDesTex05}, where the concept is also discussed for continuous models with unbounded spectrum).
For example, in the trivial case $N=1$ (one mass pinned and one spring), comparison of the form
$x_2(\lambda)=(\omega_1^2-\lambda)\,m_1/\K_1$ with the solution of the recurrence given above, $x_2(\lambda) = 1 -\lambda\,m_1/\K_1$, shows that the eigenfrequency is $\omega_1=\sqrt{\K_1/m_1}$, as it should.

To make connection with the spectral information and the localization properties, we introduce the \textit{complex Lyapunov exponent} (or ``characteristic exponent'')  \cite{Nie82,Luc92,ComTexTou13}~:
\begin{equation}
  \label{eq:DefOmega}
  \Omega(\lambda) \eqdef \lim_{N\to\infty} \frac{1}{N} \ln( x_{N+1}(\lambda)  )
\end{equation}
which is well defined when the spectral parameter $\lambda$ is not in the spectrum, i.e. for $\lambda<\omega_1^2$ (the smallest eigenvalue), or for $\lambda\in\mathbb{C}$.
From \eqref{eq:SolutionIVP}, we deduce
\begin{equation}
\frac{\ln( x_{N+1}(\lambda)  ) }{N} 
=  \frac{1}{N} \sum_{j=1}^N \ln(m_j/ \K_j) + \frac{1}{N} \sum_{\alpha=1}^N \ln(\omega_\alpha^2-\lambda)
\:.
\end{equation}
Taking the limit $N\to\infty$,  we get
\begin{equation}
  \label{eq:OmegaBare}
   \Omega(\lambda) = \mean{\ln(m_j/ \K_j) } + \int\D \lambda'\, \rho(\lambda')\, \ln(\lambda'-\lambda) 
\end{equation}
where $\rho(\lambda)$ is the spectral density (density of eigenmodes per site).
Performing $\lambda\to\lambda+\I0^+$, with now $\lambda\in\mathbb{R}$, we eventually get 
\begin{equation}
  \label{eq:OmegaAndDos}
   \Omega(\lambda+\I0^+) = \gamma(\lambda) -\I\pi \, \IDoS(\lambda)  
\end{equation}
where the real part
\begin{align}
  \gamma(\lambda) &\eqdef \lim_{N\to\infty} \frac{1}{N} \ln| x_{N+1}(\lambda) |
  \\
  &= \mean{\ln(m_j/ \K_j) } + \int\D \lambda'\, \rho(\lambda')\, \ln|\lambda'-\lambda|
\end{align}
is the Lyapunov exponent and the imaginary part
\begin{equation}
\IDoS(\lambda) = \int_0^\lambda\D \lambda'\, \rho(\lambda')
\end{equation}
 is the integrated density of states (IDoS)~; for the spring chain, the lowest eigenvalue is $\omega_1^2\to0$ in the thermodynamic limit $N\to\infty$.
The Lyapunov exponent is the exponential growth rate of the solution of the initial value problem~:
following Borland's conjecture \cite{Bor63}, we assume that it measures the localization of the eigenmodes of frequency $\omega$ (solving the spectral problem), leading to define the localization length as
\begin{equation}
\xi_\omega\eqdef1/\gamma(\omega^2)
\:.
\end{equation}

\section{Duality in the spring chain}
\label{sec:Duality}

Duality of disordered bosonic systems has been discussed in several papers, e.g. see \cite{GurCha03}, 
and originates from the symmetry between coordinates $\mathbf{x}$ and conjugate moments $\mathbf{p}$ in the Hamiltonian~:
\begin{equation}
  \mathcal{H}(\mathbf{p},\mathbf{x})
  = \frac{1}{2}
  \begin{pmatrix}
  \mathbf{p} & \mathbf{x}
  \end{pmatrix}
  \begin{pmatrix}
     \mathcal{M}^{-1}       & \mathcal{C} \\
     \mathcal{C}^\mathrm{T} & \mathcal{K}  
  \end{pmatrix}
  \begin{pmatrix}
  \mathbf{p} \\ \mathbf{x}
  \end{pmatrix}
\end{equation}
For the 1D model of interest here, $\mathcal{C}=0$ and the two matrices $\mathcal{M}^{-1}$ and $\mathcal{K}$ have very different structures~: 
the mass matrix $\mathcal{M}$ is diagonal and the spring constant matrix $\mathcal{K}$ is tridiagonal.
Hence duality is not that obvious at this level.
Still, the random matrix product formulation allows to make a precise statement. 
Introducing 
\begin{equation}
  \label{eq:DefYn}
  y_n =  K_{n-1}\, ( x_{n} - x_{n-1} )
\end{equation}
we can rewrite the recurrence \eqref{eq:TheRecurrence} in the form 
\begin{equation}
  \begin{cases}
    \displaystyle
     x_{n+1} = \left( 1 - \frac{m_n\lambda}{\K_n} \right) x_n + \frac{y_n}{\K_n}
     \\
    \displaystyle
     y_{n+1} = - m_n\lambda\,x_n + y_n
  \end{cases}
\end{equation}
which means that the vector $(x_n\,,\,y_n)$ evolves under the action of the matrix
\begin{equation}
  \label{eq:RandomMatrices}
  M_n 
  =
  \begin{pmatrix}
   1 & \K_n^{-1} \\ 0 & 1 
  \end{pmatrix}
    \begin{pmatrix}
   1 & 0 \\ -m_n\lambda & 1 
  \end{pmatrix}
  \:.
\end{equation}
This observation shows that the problem studied here can be reformulated in terms of random matrix product 
$\Pi_N=\prod_{n=1}^NM_n$~;
the Lyapunov exponent is then defined as $\gamma=\lim_{N\to\infty}(1/N)\ln||\Pi_N||$ where the norm measures the matrix elements
 (see \cite{ComTexTou13,Tex20} for recent references). 
Interestingly, random matrices of the form \eqref{eq:RandomMatrices} have also appeared in the study of a continuous string model $-\psi''(x) + \omega^2\, m'(x)\psi(x)=0$ in Ref.~\cite{ComTou17}, where the connection to spring chains was pointed out.
The random matrix product formulation makes clear that the Lyapunov exponent is invariant under exchange  
$ 1/\K_n\leftrightarrow-m_n\,\lambda $ 
which follows from the fact that products of random i.i.d. matrices $M_n$ and products of transposed matrices $M_n^\mathrm{T}$ have the same Lyapunov exponent.

Furthermore, using 
\begin{align}
 & \begin{pmatrix}
   1 & 0 \\ -m_n\lambda & 1 
  \end{pmatrix}
  \\\nonumber 
= &
    \begin{pmatrix}
   1/\sqrt{-\lambda} & 0 \\ 0  & \sqrt{-\lambda} 
  \end{pmatrix}
    \begin{pmatrix}
   1 & 0 \\ m_n  & 1 
  \end{pmatrix}
    \begin{pmatrix}
   \sqrt{-\lambda} & 0 \\ 0  & 1/\sqrt{-\lambda} 
  \end{pmatrix}
\end{align}
shows that the product of matrices
\begin{equation}
  \label{eq:RandomMatrices2}
  \widetilde{M}_n 
  =
  \begin{pmatrix}
   1 & -\lambda\K_n^{-1} \\ 0 & 1 
  \end{pmatrix}
    \begin{pmatrix}
   1 & 0 \\ m_n & 1 
  \end{pmatrix}
\end{equation}
have the same Lyapunov exponent as the product of matrices $M_n$'s.
Hence the spectral and localization properties are invariant under exchange 
\begin{equation}
  1/\K_n \hspace{0.25cm} \leftrightarrow\hspace{0.25cm} m_n
  \:.
\end{equation}
This observation is used at several places in the paper.


\section{A combinatorial approach}
\label{sec:Combinatorics}

\subsection{The analytical solution of the initial value problem}

As explained above, the solution $x_{n+1}(\lambda)$ of the initial value problem for $x_1=1$ can be written under the form of a polynomial of degree $n$ in the spectral parameter~$\lambda$ 
\begin{equation}
  \label{eq:PolynomialXn}
  x_{n+1}(\lambda) = \sum_{k=0}^n a_{n+1}^{(k)}\,(-\lambda)^k
  \:.
\end{equation}
Obviously $a_{n+1}^{(0)}$ solves the same equation as $x_{n+1}(0)$, which can be conveniently written as 
\begin{equation}
  \label{eq:ReccComp0}
    a_{n+1}^{(0)} - a_{n}^{(0)} = \frac{\K_{n-1}}{\K_n} \left( a_{n}^{(0)} - a_{n-1}^{(0)} \right)
    \:.
\end{equation}
For free boundary condition we have $\K_0=0$, hence $a_{2}^{(0)} = a_{1}^{(0)}$ and therefore $a_{n}^{(0)}=1$ $\forall\,n\geq1$.
The value $a_{0}^{(k)}$ plays no role, however it may be convenient to set $a_{0}^{(k)}=0$ below.
Introducing \eqref{eq:PolynomialXn} in \eqref{eq:TheRecurrence}, we obtain the recurrence for the coefficients with $k\geq1$,
\begin{align}
  \label{eq:20}
    a_{n+1}^{(k)} - a_{n}^{(k)} = \frac{\K_{n-1}}{\K_n} \left( a_{n}^{(k)} - a_{n-1}^{(k)} \right)
     + \frac{m_{n}}{\K_n}a_{n}^{(k-1)} 
    \:,
\end{align}
with initial value $a_{1}^{(k)}=0$.
Compared to \eqref{eq:ReccComp0}, the second term in \eqref{eq:20} plays the role of a source term. 
It is convenient to introduce $u_{n+1}^{(k)} = a_{n+1}^{(k)} - a_{n}^{(k)}$, i.e.
$a_{n+1}^{(k)}=\sum_{m=0}^nu_{m+1}^{(k)}$.
Thus, $a_{0}^{(0)}=0$ and $a_{n}^{(0)}=1$ for $n\geq1$ corresponds to $u_{n}^{(0)}=\delta_{n,1}$.
For $k>0$, the recurrence takes the form
\begin{equation}
  u_{n+1}^{(k)}  = \frac{\K_{n-1}}{\K_n} \, u_{n}^{(k)}  + \frac{m_{n}}{\K_n} \sum_{j=1}^n u_{j}^{(k-1)}
  \:.
\end{equation}
More clearly, we can introduce $b_{n}^{(k)}=\K_{n-1}\,u_{n}^{(k)}$ for $n>1$, which is the coefficient of the expansion of \eqref{eq:DefYn} in powers of $\lambda$~: the recurrence \eqref{eq:20} is a simple arithmetic series $b_{n+1}^{(k)}  = b_{n}^{(k)}  +  m_{n}\,a_{n}^{(k-1)} $.
The solution is 
\begin{equation}
  \label{eq:IterationU0}
  u_{n+1}^{(k)}  
  = \frac{1}{\K_n}  \sum_{j=1}^n m_j \sum_{p=1}^j u_{p}^{(k-1)}  
  = \frac{1}{\K_n}  \sum_{p=1}^n u_{p}^{(k-1)} \sum_{j=p}^n m_j 
  \:.
\end{equation}
We now remark that the first terms of the sum $\sum_p$ vanish since $u_p^{(k-1)}=0$ for $p<k$, hence we obtain the form
\begin{equation}
  \label{eq:IterationU}
    u_{n+1}^{(k)}  =  \frac{1}{\K_n}\sum_{p=k-1}^{n-1}  u_{p+1}^{(k-1)}\,\sum_{j=p+1}^{n} m_j 
    \:,
\end{equation}
 now appropriate for iteration.
 The first iterations give
\begin{align}
   u_{n+1}^{(1)}  &=  \frac{1}{\K_n} \sum_{j=1}^{n} m_j 
   \:,
   \\
   u_{n+1}^{(2)}  &=   \sum_{p=1}^{n-1}  
    \frac{1}{K_{p}} \sum_{j=1}^{p} m_j \,
    \frac{1}{\K_n} \sum_{j'=p+1}^{n} m_{j'} 
    \nonumber
    \\
  &=  
  \sum_{1\leq j_1\leq i_1<j_2\leq n}
  \frac{m_{j_1}m_{j_2}}{\K_{i_1}\K_{n}}
      \:,
 \end{align} 
 etc.
We deduce $a_{n+1}^{(k)}=\sum_{p=k}^n u_{p+1}^{(k)}$~:
\begin{equation}
  \label{eq:MainRes}
  a_{n+1}^{(k)} 
  = 
  \hspace{-0.5cm}
  \sum_{1\leq j_1 \leq i_1< j_2 \leq i_2<\cdots < j_k\leq i_k\leq n}
  \   \prod_{m=1}^k \frac{m_{j_m}}{K_{i_m}}
  \:,
\end{equation}
which is Eq.~\eqref{eq:MainResIntro2}. 
It will be useful for the following to organize the sums as 
\begin{equation}
  \label{eq:MainRes2}
  a_{n+1}^{(k)} 
  = 
  \hspace{-0.5cm}
  \sum_{1\leq i_1<\cdots<i_{k}\leq n}
   \ \prod_{m=1}^k 
   \bigg(\frac{1}{\K_{i_{m}}}\sum_{j=i_{m-1}+1}^{i_{m}} m_j \bigg)
\end{equation}
with $i_0=0$.
Similarly,
\begin{equation}
  \label{eq:MainRes1}
  a_{n+1}^{(k)} = 
  \hspace{-0.5cm}
  \sum_{1\leq j_1 <\cdots<j_{k}\leq n}  
  \ \prod_{m=1}^{k} \bigg( m_{j_{m}}  
  \sum_{i=j_{m}}^{j_{m+1}-1} \frac{1}{\K_i}\bigg)
\end{equation}
with $j_{k+1}=n$.
The coefficients are also those of the spectral determinant~\eqref{eq:SpectralDet-Neumann/Dirichlet} for the tridiagonal matrix $\Lambda$~; these expressions exhibit the symmetry between masses $m_j$'s and inverse spring constants $1/\K_i$'s discussed in the previous section.
The case of a chain pinned at the two sides 
(Dirichlet/Dirichlet boundary conditions)
is also discussed in Appendix~\ref{app:DirDir}.

So far, all is exact and we did not make any assumption on the masses and the spring constants, which can be deterministic or random.

\subsection{Symmetrization of the coefficients $a_{n+1}^{(k)}$}

The starting point of our asymptotic analysis of the coefficients is to remark that when averaging any function of $n$ i.i.d. random variables, it is possible to perform any permutation of these variables.
As a consequence we can write
\begin{align}
   \mean{ F( m_1,\cdots,m_n  ) }   
   &= \mean{ F( m_{\pi(1)},\cdots,m_{\pi(n)}  )   }
   \\\nonumber
   &= \mean{ \ \symmetriz{ F( m_{\pi(1)},\cdots,m_{\pi(n)}  ) }_{\pi\in\mathcal{S}_n} }
 \end{align} 
where $\pi$ is a permutation and $\symmetriz{ \cdots }_{\pi\in\mathcal{S}_n}$ denotes full symmetrization, i.e. averaging over the symmetric group $\mathcal{S}_n$ with uniform weight.
Here, we have two sets of i.i.d. random variables, $\{m_i\}$ and $\{\K_i\}$, hence we symmetrize with respect to permutations of both sets.
This observation leads to the idea to consider the coefficients symmetrized with respect to the permutations of random variables, \textit{without} disorder average.
We assume that this will provide the typical behaviour of these coefficients, which will indeed be verified on few well controlled specific cases.
We consider 
\begin{equation}
  \symmetriz{ a_{n+1}^{(k)} }_{\pi,\sigma} = 
  \symmetriz{ 
  \sum_{1\leq j_1\leq i_1 \cdots }\  \prod_{m=1}^k \frac{m_{\pi(j_m)}}{K_{\sigma(i_m)}}  
  }_{\pi,\:\sigma\in\mathcal{S}_n}
  \:.
\end{equation}
Noticing that 
\begin{equation}
  \symmetriz{ \prod_{m=1}^k m_{\pi(j_m)}  }_{\pi\in\mathcal{S}_n} 
  = \symmetriz{ \prod_{m=1}^k m_{\pi(m)}  }_{\pi\in\mathcal{S}_n} 
\end{equation}
allows to remove the dependence of the masses and spring constants in the indices of the $2k$ sums. 
This remark greatly simplifies the multiple sums~:
\begin{align}
  \label{eq:SymmetrizedCoeff}
  \symmetriz{ a_{n+1}^{(k)} }_{\pi,\sigma} = &
  \symmetriz{ \prod_{j=1}^k m_{\pi(j)}  }_{\pi\in\mathcal{S}_n} 
  \symmetriz{ \prod_{i=1}^k \K^{-1}_{\sigma(i)}  }_{\sigma\in\mathcal{S}_n} 
  \nonumber
  \\  
  &\hspace{1cm}\times
  \hspace{-0.15cm}
  \underbrace{ 
  \sum_{1\leq j_1 \leq i_1< j_2 \leq i_2<\cdots < j_k\leq i_k\leq n} 
 \hspace{-1.55cm} 
 1
 \hspace{1.25cm}
  }_{=S_k(n)}            
\end{align}
where the sum $S_k(n)$ is computed in Appendix~\ref{app:Sums}~:
\begin{equation}
  \label{eq:DefSnk}
  S_k(n) = 
  \hspace{-0.5cm}  
  \sum_{1 \leq i_1<i_2 < \cdots < i_k \leq n}\: \prod_{m=1}^{k} (i_{m}-i_{m-1})
  =
  \begin{pmatrix}
    n+k \\ 2k
  \end{pmatrix}
\end{equation}
(with $i_0=0$). 
In the next subsection, we analyze the remaining symmetrized products.

\subsection{Symmetrized product of $k$ random variables among $n$~: a fermionic random energy model}
\label{subsec:SymmetrizedProducts}

Let us study the symmetrized product of $k$ i.i.d. random variables chosen among $n$ ones, $X_1,\ldots,X_n$ (the variables correspond either to the inverse spring constants $K_j^{-1}$'s of the masses $m_j$'s).
We have
\begin{equation}
  \label{eq:ExpressionOfProductOfkVariables}
  \symmetriz{ \prod_{j=1}^k X_{\pi(j)} }_{\pi\in\mathcal{S}_n} 
  = 
  \frac{k!(n-k)!}{n!}\, Z(k,n) 
\end{equation}
where 
\begin{equation}
  \label{eq:ZforOrderedIndices}
  Z(k,n) = 
  \sum_{1\leq i_1 < i_2 <\cdots < i_k\leq n}  \ \prod_{m=1}^k X_{i_m}
  \:.
\end{equation}
It is helpful to use the analogy with the canonical partition function for $k$ non-interacting fermions occupying $n$ random energy levels, which is clear if we write $X_i=\EXP{-\beta\varepsilon_i}$, where $\varepsilon_i$ is a single particle energy level and $\beta$ the inverse temperature.
Introducing the occupation number $n_i\in\{0,\,1\}$ of level  $\varepsilon_i$ allows to rewrite more conveniently the canonical partition function as 
\begin{equation}
  Z(k,n) = 
  \sum_{  \underset{\sum_in_i=k}{n_1,\cdots,n_n}  }
  \prod_{i=1}^n X_{i}^{n_i}
  \:.
\end{equation}
This form makes clear that $Z(k,n) = n!/\big[k!(n-k)!\big]$ when $X_i=1$, which simplifies with the factor in \eqref{eq:ExpressionOfProductOfkVariables}, as it should in this trivial case.
It is convenient to introduce the grand partition function
\begin{equation}
  \Xi(\varphi,n) = \sum_{k=0}^n \varphi^k\, Z(k,n) 
  = \prod_{i=1}^n \left(1 + \varphi\,X_{i} \right)
  \:,
\end{equation}
where $\varphi$ is the fugacity \cite{TexRou24book}.
We can go back to the canonical partition function thanks to 
\begin{equation}
  \label{eq:FromXitoZ}
  Z(k,n) 
  = \oint \frac{\D\varphi}{2\I\pi} \, \frac{\Xi(\varphi,n) }{\varphi^{k+1}}
  \:,
\end{equation}
where the contour encircles the origin once in the complex plane of $\varphi$.
In the limit $k\gg1$, we expect that the integral \eqref{eq:FromXitoZ} is dominated by a saddle point $\varphi_*$ solving
\begin{equation}
  \label{eq:SaddleForZ}
  \derivp{}{\varphi}
    \left[
      \ln \Xi(\varphi,n) - k \,\ln\varphi
    \right]
  \Big|_{\varphi=\varphi_*}
  =0
  \:.
\end{equation}
This equation has a clear physical meaning within the fermionic model~:
$\overline{k}^\mathrm{g}=\partial \ln \Xi(\varphi,n)/\partial\ln\varphi$ is the (grand canonical) average number of fermions,
hence the saddle point equation \eqref{eq:SaddleForZ} corresponds to express the fugacity as a function of the number of fermions 
\begin{equation}
  \label{eq:NumberOfFermions}
  k = \sum_{i=1}^n \frac{1}{1/(X_i\varphi_*)+1}
  \:,
\end{equation}
where the function in the sum is the Fermi-Dirac distribution. 
In other terms, this equation gives the canonical fugacity (fugacity as a function of the number of fermions), denoted $\varphi_*(k/n)$.
Then we deduce 
$  
\ln  Z(k,n) \simeq \ln \Xi(\varphi_*,n) - k \,\ln\varphi_*
$.
Let us now take advantage of the extensivity property, in terms of the number $n$ of levels.
In the limit of large $n$, assuming $\smean{(\ln X)^2}<\infty$, we have
\begin{equation}
  \ln \Xi(\varphi,n)  
  = \sum_{i=1}^n \ln\left(1 + \varphi\,X_{i} \right)
  = n\,\mean{ \ln(1 + \varphi\,X_i) } + \mathcal{O}(\sqrt{n})
\end{equation}
We can write a similar relation for the number of fermions, hence introducing the fraction of occupied levels 
\begin{equation}
   \theta = \frac{k}{n}
   \:,
\end{equation}
in the limit $n\to\infty$, Eq.~\eqref{eq:NumberOfFermions} rewrites
\begin{equation}
  \label{eq:SaddleForTheta}  
 \theta = \varphi_*\,\mean{ \frac{X_i}{1+\varphi_*\,X_i} }
 \in[0,1]
 \:.
\end{equation}
The $n\to\infty$ limit of \eqref{eq:NumberOfFermions} always exists thanks to the inequality $0\leq X\varphi/\big(1+X\varphi\big)\leq1$.   
This transcendental equation for $\varphi_*(\theta)$ will play a central role in the analysis. 
The $\theta\to0^+$ limit corresponds to $\varphi_*\to0$, while the $\theta\to1^-$ limit corresponds to $\varphi_*\to\infty$.
Given the fugacity, we deduce the free energy per level
\begin{equation}
  \label{eq:FreeEnergy}
  \frac{\ln Z(k,n)}{n} \simeq \mean{ \ln(1 + \varphi_* \,X_i) } - \theta\,\ln\varphi_* 
  \:.
\end{equation}
This estimation of $Z(k,n)$ thus gives the typical value (in the Physics sense) of the symmetrized product \eqref{eq:ExpressionOfProductOfkVariables}.

If we introduce the free energy per level $f(\theta)=-\ln Z(k,n)/n$, we have the ``thermodynamic identity'' $\partial f(\theta)/\partial\theta = \ln\varphi_*(\theta)$ (canonical chemical potential) \cite{TexRou24book}~;
the relation can be proven by differentiating \eqref{eq:FreeEnergy} with respect to $\theta$.
Noticing that $\ln\big[Z(0,n)\big]=0$, we can integrate this equation~:
\begin{equation}
  \label{eq:Useful1}
  \frac{\ln Z(k,n)}{n} \simeq - \int_0^{k/n} \D t\,   \ln\varphi_*(t)
  \:.
\end{equation}
Obviously $(1/n)\ln Z(n,n)=\smean{\ln X}+ \mathcal{O}(1/\sqrt{n})$, hence we have also 
\begin{equation}
  \label{eq:Useful2}
  \frac{\ln Z(k,n)}{n} \simeq  \smean{\ln X} + \int_{k/n}^1 \D t\,  \ln\varphi_*(t)
  \:.
\end{equation}
It will be more convenient to use \eqref{eq:Useful1} and \eqref{eq:Useful2} rather than \eqref{eq:FreeEnergy}.

\subsubsection{Case $\mean{X}<\infty$ and $\mean{1/X}<\infty$}

We now apply these expressions to obtain the asymptotic of $Z(k,n)$ for $n\gg k\gg1$ and for $n\gg n-k\gg 1$.
The first case requires to determine the $\theta\to0$ behaviour of the fugacity. 
Expanding \eqref{eq:SaddleForTheta} for $\varphi_*\to0$ we get 
\begin{equation}
  \label{eq:FugacitySmallNbStandardCase}
  \varphi_*(\theta) \simeq \frac{\theta}{\mean{X}} + \frac{\smean{X^2}\theta^2}{\mean{X}^3}
  \hspace{0.5cm}\mbox{for }\theta\to0
  \:. 
\end{equation}
From \eqref{eq:Useful1} we deduce 
\begin{equation}
  \label{eq:ExpansionFreeEnergySmallK}
  \frac{\ln Z(k,n)}{n} \simeq \theta\, \ln(\EXP{}\mean{X}/\theta)  -\theta^2\frac{\mean{X^2}}{2\mean{X}^2}
  + \mathcal{O}(\theta^3)
  \:.
\end{equation}
Adding the contribution of 
$(1/n)\ln\big[k!(n-k)!/n!\big]=\theta\ln\theta+(1-\theta)\ln(1-\theta)\simeq-\theta\,\ln(\EXP{}/\theta)+\theta^2/2$, we obtain the typical value
\begin{equation}
  \label{eq:AsympSymmProductSmallk}
  \symmetriz{ \prod_{j=1}^k X_{\pi(j)}  }_{\pi\in\mathcal{S}_n} 
   \hspace{-0.25cm}
   \sim \mean{X}^k   \exp\left\{ -\frac12 \left(\frac{\smean{X^2}}{\smean{X}^2} -1 \right) \frac{k^2}{n} \right\}
\end{equation}
for $1\ll k\ll n$. 
Clearly, the factor $\mean{X}^k$ corresponds to non random variables, while the fluctuations diminish the typical value of the product.
We will also discuss cases where $\mean{X}=\infty$ or $\mean{X^2}=\infty$ in the following.

Let us now consider $\varphi_*\to\infty$ in \eqref{eq:SaddleForTheta}~:
we can write
\begin{equation}
  1-\theta = \mean{ \frac{1}{1+\varphi_*\,X} }
\end{equation}
hence we can use the symmetry $\theta\leftrightarrow1-\theta$, $\varphi\leftrightarrow1/\varphi$ and $X\leftrightarrow1/X$ in order to relate the $\theta\to1^-$ to the $\theta\to0^+$ behaviour.
As a result 
\begin{align}
  \label{eq:ExpansionFugacityLarge}
  \frac{\ln Z(k,n)}{n} \simeq 
  &\mean{\ln X} 
  + (1-\theta) \, \ln\big[\EXP{}\mean{1/X}/(1-\theta)\big]  
  \nonumber\\
  &- (1-\theta)^2 \,\frac{\mean{1/X^2}}{2\mean{1/X}^2}
  + \mathcal{O}((1-\theta)^3)
  \:.
\end{align}
i.e.
\begin{align}
  \label{eq:AsympSymmProductLargek}
   \symmetriz{ \prod_{j=1}^k X_{\pi(j)}  }_{\pi\in\mathcal{S}_n} 
   \hspace{-0.5cm}
   &\sim \EXP{n\mean{\ln X}} \mean{1/X} ^{n-k} 
   \\\nonumber
   &\times
    \exp\left\{ -\frac12 \left(\frac{\smean{1/X^2}}{\smean{1/X}^2} -1 \right) \frac{(n-k)^2}{n} \right\}
\end{align}
for $1\ll n-k\ll n$.

\subsubsection{A case where $\mean{1/X}=\infty$}

If the distribution $q(x)$ of $X$ is such that $q(0)$ is finite, we have $\mean{1/X}=\infty$.
For example, this occurs for the exponential distribution $q(x)=a\,\EXP{-ax}$ which we will consider below. 
Then, Eq.~\eqref{eq:SaddleForTheta} takes the form 
\begin{equation}
  \label{eq:FugacityForExpVar}
  1 - \frac{a}{\varphi_*}\,\EXP{a/\varphi_*}\,E_1(a/\varphi_*) = \theta
\end{equation}
where $E_1(z)=\int_z^\infty(\D t/t)\,\EXP{-t}$ is the exponential integral \cite{DLMF}.
An expansion for $\varphi_*\to0$ leads to (\ref{eq:FugacitySmallNbStandardCase},\ref{eq:ExpansionFreeEnergySmallK}), as it should.

When $\varphi_*\to\infty$, we obtain 
$\varphi_*(\theta)\simeq a\, (1-\theta)^{-1}\ln\big[\EXP{-\mathbf{C}}/(1-\theta)\big]$ for $\theta\to1$, where $\mathbf{C}\simeq0.577216$ is the Euler-Mascheroni constant.
The behaviour obtained above for $\mean{1/X}<\infty$, receives an additional logarithmic correction. 
Using \eqref{eq:Useful2} we get
\begin{equation}
  \label{eq:ProdForInfiniteMeanInvX}
   \symmetriz{ \prod_{j=1}^k X_{\pi(j)}  }_{\pi\in\mathcal{S}_n} 
   \hspace{-0.5cm}
   \sim \EXP{n\mean{\ln X}}
  \left[
    a\, 
    \ln\left(\frac{n\, \EXP{-\mathbf{C}}}{n-k}\right)
  \right]^{n-k}
\end{equation}
for $1\ll n-k\ll n$. Here $\mean{\ln X}=-\ln a -\mathbf{C}$.

\subsubsection{Power-law distribution~: a case with $\mean{X}=\infty$}

In the paper we are mostly interested in distributions with power-law tails, like $q(x)=\mu\,x^{-1-\mu}$ for $x>1$, such that the mean value might be divergent (when $\mu\in]0,1]$).
For this distribution, the equation \eqref{eq:SaddleForTheta} takes the form 
\begin{equation}
  \label{eq:SaddleForThetaPowerLaw}
  _2F_1\left( 1 , \mu ; 1+\mu ; -1/\varphi_* \right) = \theta
\end{equation}
where $_2F_1( a , b ; c ; z )$ is the hypergeometric function \cite{gragra,DLMF}.
As a check, we first consider $\theta\to1^-$, then we can use $_2F_1( a , b ; c ; z )\simeq 1 + a\,b\,z/c$ for $z\to0$, leading to $\varphi_*(\theta)\simeq\big[\mu/(1+\mu)\big]/(1-\theta)$.
We recognize $\mean{1/X}=\mu/(1+\mu)$, hence \eqref{eq:AsympSymmProductLargek} holds with $\mean{\ln X}=1/\mu$.

Let us now discuss the behaviour for $\theta\to0^+$. 
For this purpose we use a well-known functional relation for the hypergeometric function 
\begin{align}
  &_2F_1\left( 1 , \mu ; 1+\mu ; -1/\varphi \right)
  \\\nonumber
   &= \Gamma(1+\mu)\Gamma(1-\mu)\,\varphi^\mu +\frac{\mu\,\varphi}{\mu-1}\,_2F_1\left( 1 , 1-\mu ; 2-\mu ; -\varphi \right) 
\end{align}
Hence the expansion for $\varphi\to0^+$ has the form of an analytic series plus the non analytic term $\mathcal{O}(\varphi^\mu)$~:
\begin{align}
  \label{eq:Expansion2F1}  
  &_2F_1\left( 1 , \mu ; 1+\mu ; -1/\varphi \right)
  \\\nonumber
  \simeq &
  \frac{\pi\mu}{\sin\pi\mu}\,\varphi^\mu
  + \frac{\mu}{\mu-1} \,\varphi 
  - \frac{\mu}{\mu-2} \,\varphi^2
  + \mathcal{O}(\varphi^3)
  \:.
\end{align}
Depending on $\mu$, the expansion is dominated by the $\mathcal{O}(\varphi)$ or the $\mathcal{O}(\varphi^\mu)$ term.
As a result, the possible limiting behaviours of the solution of \eqref{eq:SaddleForThetaPowerLaw} for $\varphi\to0$, i.e. for $\theta\to0$, are
\begin{align}
  \label{eq:FugacityForSmallTheta}
  \varphi_*(\theta)
  \underset{\theta\to0}{\simeq}
  \begin{cases}
    \Big( \frac{\sin\pi\mu}{\pi\mu} \, \theta \Big)^{1/\mu}  & \mbox{for } 0<\mu<1 
    \\[0.125cm]
    \frac{\theta}{\ln(1/\theta)}                                & \mbox{for } \mu=1
    \\[0.125cm]
    \frac{\theta}{\mean{X}} - \frac{\pi\mu}{\sin\pi\mu}\, \frac{\theta^{\mu}}{\mean{X}^{\mu+1}}
                                                                & \mbox{for } 1<\mu<2
    \\[0.125cm]
    \frac{\theta}{\mean{X}} + \frac{\theta^2}{4}\,\ln(2/\theta) & \mbox{for } \mu=2   
    \\[0.125cm]
    \frac{\theta}{\mean{X}} +  \frac{\mean{X^2}\theta^2}{\mean{X}^3}      & \mbox{for } \mu>2  
  \end{cases}
\end{align}
with $\mean{X^k}=\mu/(\mu-k)$ for $k<\mu$. 
Let us focus on the case $\mu<1$~:
we get from \eqref{eq:Useful1} the behaviour
\begin{align}
  \label{eq:ZknForInfiniteMeanX}
  Z(k,n) \sim  
  \left( \frac{\pi\mu}{\sin\pi\mu} \, \frac{n\,\EXP{}}{k} \right)^{k/\mu}
  \hspace{0.5cm}\mbox{for }
  1\ll k\ll n
\end{align}
and 
\begin{equation}
  \label{eq:PRVForInfiniteMeanX}
   \symmetriz{ \prod_{i=1}^k X_{\pi(i)}  }_{\pi\in\mathcal{S}_n} 
  \sim \left( \frac{\pi\mu}{\sin\pi\mu}  \right)^{k/\mu} 
  \left( \, \frac{n\,\EXP{}}{k} \right)^{k(1-\mu)/\mu}
  \:.
\end{equation}

Let us also discuss the marginal case $\mu=1$, which is a bit more tricky. Eq.~\eqref{eq:SaddleForThetaPowerLaw} takes the simple form
\begin{equation}
  \varphi_*\,\ln ( 1 + 1/\varphi_* ) = \theta
  \:.
\end{equation}
Using $\mean{\ln(1+\varphi\,X)}=(1+\varphi)\,\ln(1+\varphi)-\varphi\,\ln\varphi$ we obtain
\begin{align}
  \frac{ \ln Z(k,n) }{ n } \simeq \left( \frac{1}{\varphi_*} + 1 - \ln \varphi_* \right)\theta + \ln \varphi_* 
\end{align}
leading to 
\begin{equation}
  \label{eq:ZknForMarginalCase}
   Z(k,n) \sim \left(  \frac{n\,\EXP{}}{k}\,\ln (n/k)  \right)^{k}
\end{equation}
i.e.
\begin{equation}
  \label{eq:PRVForMarginalCase}
  \symmetriz{ \prod_{i=1}^k X_{\pi(i)}  }_{\pi\in\mathcal{S}_n} 
  \sim \left( \ln \frac nk \right)^k
  \hspace{0.5cm}\mbox{for }
  1\ll k\ll n
  \:.
\end{equation}

\subsection{Asymptotics of the coefficients $a_{n+1}^{(k)}$ and the complex Lyapunov exponent}

Using the decomposition \eqref{eq:SymmetrizedCoeff}, we can now study the precise asymptotic of the coefficients $a_{n+1}^{(k)}$ from which we deduce $\Omega$.
This will lead straightforwardly to the complex Lyapunov exponent~: 
we identify the optimal value of the index $k$ 
which solves
\begin{equation}
  \label{eq:SaddleForK}
  \deriv{}{k} \big\{ \ln a_{n+1}^{(k)} + k \ln(-\lambda) \big\}\Big|_{k=k_*}=0
\end{equation}
and dominates the sum \eqref{eq:MainRes0}~:
$x_{n+1}(\lambda)\sim a^{(k_*)}_{n+1}\,(-\lambda)^{k_*}$.
Using the bounds $a^{(k_*)}_{n+1}\,(-\lambda)^{k_*}\leq x_{n+1}(\lambda)\leq n\,a^{(k_*)}_{n+1}\,(-\lambda)^{k_*} $ we deduce the complex Lyapunov exponent 
\begin{equation}
  \label{eq:LyapFromOptimum}
  \Omega(\lambda) = \lim_{n\to\infty}\frac{\ln \big( a^{(k_*)}_{n+1}\,(-\lambda)^{k_*} \big)}{n}
  \:.
\end{equation}
We will see below that this approach provides the asymptotic behaviours (for small $\lambda$ or large $\lambda$) in a straightforward manner.
From the series representation \eqref{eq:MainRes0}, it is clear that the $\lambda\to0^-$ limit involves coefficients 
for $k\ll n$,
while the $\lambda\to-\infty$ limit is related to large $k$, close to $n$.

Before entering into the detailed analysis, let us make a final remark~:
the complex Lyapunov exponent \eqref{eq:DefOmega} characterizes the typical behaviour of the solution $x_{n+1}(\lambda)$ of the initial value problem, i.e. the average of $\ln x_{n+1}(\lambda)$.
Similarly, \eqref{eq:LyapFromOptimum} shows that we only need to determine the typical value of the coefficients $a^{(k)}_{n+1}$, which justifies the study of the typical value of the random product of the previous subsection.

Now consider 
\begin{align}
  \label{eq:AsymptoticOfAnp1k}
  &\frac{\ln \big( \ssymmetriz{a^{(k)}_{n+1}}_{\pi,\sigma}\,(-\lambda)^{k} \big)}{n}
  = 
  \frac{\ln S_k(n)+2\ln\big[k!(n-k)!/n!\big]}{n} 
  \nonumber\\
  &\hspace{1cm}+ \frac{k}{n}\ln(-\lambda) + \frac{\ln Z^{(1/\K)}(k,n)}{n} + \frac{\ln Z^{(m)}(k,n)}{n}
  \nonumber\\
  & 
  \underset{n\to\infty}{=}
   (1+\theta)\ln(1+\theta)+(1-\theta)\ln(1-\theta)+\theta\ln\big[(-\lambda)/4\big]
  \nonumber\\
  &\hspace{1cm}
   -\int_0^\theta\D t\, \ln \varphi_*^{(1/\K)}(t)-\int_0^\theta\D t\, \ln\varphi_*^{(m)}(t)  
\end{align}
where $\theta=k/n$. The partition functions $Z^{(1/\K)}(k,n)$ and $Z^{(m)}(k,n)$ are related to the symmetrized products of inverse spring constants and masses, respectively, and $\varphi_*^{(1/\K)}(\theta)$, $\varphi_*^{(m)}(\theta)$ the two corresponding fugacities, solutions of the transcendental equations (\ref{eq:Transcendental-K},\ref{eq:Transcendental-m}).
Removing $(-\lambda)^k$ in Eq.~\eqref{eq:AsymptoticOfAnp1k} gives the asymptotic of the symmetrized coefficient $a^{(k)}_{n+1}$.
Assuming that we can replace the coefficient in \eqref{eq:LyapFromOptimum} by the symmetrized one, we get
\begin{align}
  \label{eq:MainResultForOmega}
  \Omega(\lambda) = - \underset{\theta}{\mathrm{min}}
  \int_0^\theta\D t\, \ln\left[
    \frac{4}{(-\lambda)}\,\frac{1-t}{1+t}\,\varphi_*^{(1/\K)}(t)\varphi_*^{(m)}(t)
  \right]
  \:.
\end{align}
The position of the minimum, which we will denote $\theta_*(\lambda)=k_*/n\in[0,1]$, is given by solving
\begin{equation}
  \label{eq:ThetaStar}
  \varphi_*^{(1/\K)}(\theta)\varphi_*^{(m)}(\theta) = \frac{(-\lambda)}{4}\,\frac{1+\theta}{1-\theta}
  \:,
\end{equation}
which is equivalent to \eqref{eq:SaddleForK}.
Eq.~\eqref{eq:MainResultForOmega} is a central result.
We stress that the only approximation which has been used to get \eqref{eq:MainResultForOmega} is the symmetrization of the coefficients, $a^{(k)}_{n+1}\simeq\ssymmetriz{a^{(k)}_{n+1}}_{\pi,\sigma}$.
From this respect, \eqref{eq:MainResultForOmega} is not an exact formula, although we will see that it provides accurate predictions.

Expressing $Z(k,n)$ with \eqref{eq:Useful2}, we get the alternative representation
\begin{align}
  \label{eq:MainResultForOmegaBis}
  \Omega(\lambda) 
  &=
  \ln(-\lambda) + \mean{\ln(m/K)}
  \\\nonumber
  &+
   \underset{\theta}{\mathrm{max}}
  \int_\theta^1\D t\, \ln\left[
    \frac{4}{(-\lambda)}\,\frac{1-t}{1+t}\,\varphi_*^{(1/\K)}(t)\varphi_*^{(m)}(t)
  \right]
  \:,
\end{align}
which will be more appropriate for the study of the $\lambda\to-\infty$ limit.

\subsection{Low and high frequency expansions when $\mean{m_i}$, $\mean{\K_i}$, $\mean{1/m_i}$ and $\mean{1/\K_i}$ are finite}

Let us first apply the formula to the simple case where 
\begin{align*}
 \mb = \mean{m_j} <\infty
 \hspace{0.5cm}\mbox{and}\hspace{0.5cm}
 1/\Kb = \mean{1/ \K_j} <\infty
 \:,
\end{align*}
and consider the $\lambda\to0^-$ limit.
Then the solution of \eqref{eq:ThetaStar} is small $\theta_*=k_*/n\ll1$, hence we can simply keep the leading term in Eq.~\eqref{eq:FugacitySmallNbStandardCase}~: $\varphi^{(1/\K)}(\theta)\simeq\Kb\theta$ and $\varphi^{(m)}(\theta)\simeq\theta/\mb$. 
Therefore \eqref{eq:ThetaStar} has solution
\begin{equation}
  \label{eq:ThetaStarNormalCaseSmallLambda}
  \theta_*(\lambda) \simeq \frac12\sqrt{-\lambda\mb/\Kb }
  \:.
\end{equation}
The integral \eqref{eq:MainResultForOmega} is then straightforward
\begin{align}
  \Omega 
  &\simeq 
  -\int_0^{\theta_*} \hspace{-0.25cm}
  \D t \,   
  \ln\left[
    \frac{4}{(-\lambda)}\,\frac{1-t}{1+t}\,\frac{\Kb\,t^2}{\mb}
  \right]
  \simeq \theta_* \,  \ln\left[
    \frac{\mb(-\lambda)}{\Kb}\,\frac{\EXP{2}}{4\theta_*^2} 
  \right]
\end{align}
Equivalently, using $\theta_*=k_*/n$ and removing the factor $(-\lambda)^{k_*}$, this corresponds to the asymptotic of the symmetrized coefficients
\begin{equation}
  \label{eq:MeanAknStandardCase}
  \symmetriz{ a_{n+1}^{(k)} }_{\pi,\sigma}
  \sim \left( \frac{\mb}{\Kb } \right)^k \left(\frac{n\,\EXP{}}{2k}\right)^{2k}
   \hspace{0.5cm}
   \mbox{for }
   1\ll k\ll n  
  \:.
\end{equation}
%
After simplification, we eventually get $\Omega \simeq 2\theta_*$, i.e.
\begin{equation}
  \label{eq:Omega-SmallLambda}
  \Omega(\lambda) \simeq 
  \sqrt{-\frac{\mb\lambda}{\Kb }}
    \hspace{0.5cm}
  \mbox{for }
  \lambda\to0^-
  \:.
\end{equation}
This result was obtained by making use of the simplest linear approximation for the fugacities.
However, they can be determined more accurately under the form of a systematic series in the occupation $\theta$, which provides an expansion of $\Omega$ in powers of $\sqrt{-\lambda}$.
We will apply this below.

\begin{figure}[!h]
\centering
\includegraphics[width=0.4\textwidth]{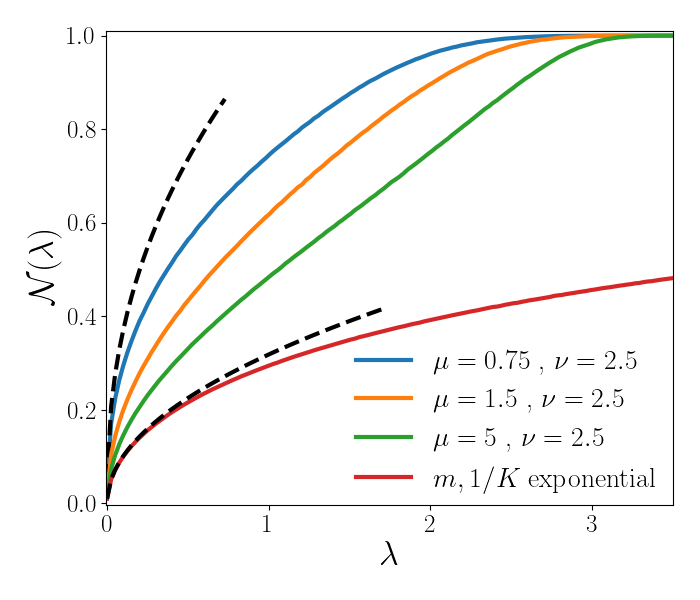}
\caption{\it IDoS obtained numerically.
 In orange~: 
  the case of exponential distributions (with $\mean{\K}=\mean{1/m}=\infty$) shows a slow convergence towards unity at high frequency.  
 In blue~: 
 the IDoS for power-law distribution (with $\mean{\K}<\infty$ and $\mean{1/m}<\infty$) for $\mu=3/4$ and $\nu=5/2$ exhibits a rapid convergence (Lifshitz tail). 
 Black dashed lines are our analytical asymptotic predictions (no adjustable parameter).}
\label{fig:IDoS}
\end{figure}

We now consider the limit $\lambda\to-\infty$,
assuming 
\begin{align*}
 1/\tilde{m} = \mean{1/m_j} <\infty
 \hspace{0.5cm}\mbox{and}\hspace{0.5cm}
 \overline{\K} = \mean{\K_j} <\infty
 \:.
\end{align*}
In this case the solution of \eqref{eq:ThetaStar} is large $\theta_*=k_*/n\sim1$, hence we use $\varphi_*^{(1/\K)}(\theta)\simeq\overline{K}/(1-\theta)$ and $\varphi_*^{(m)}(\theta)\simeq1/\big[\tilde{m}\,(1-\theta)\big]$, leading to 
\begin{equation}
  \theta_*(\lambda) \simeq 1 - \frac{2\overline{\K}}{\tilde{m}\,(-\lambda)}
  \:.
\end{equation}
From \eqref{eq:MainResultForOmegaBis} we get 
\begin{align}
  \label{eq:ProvExpForOmega}
  \Omega 
  &\simeq 
  \ln(-\lambda) + \mean{\ln(m/\K)}
  +\int_{\theta_*}^1 \D t\,
  \ln\left[  
    \frac{4\overline{\K}}{\tilde{m}\,(-\lambda)(1-t^2)}
  \right]
  \nonumber
  \\
  &\simeq
  \ln(-\lambda) + \mean{\ln(m/\K)}
  + (1-\theta_*)\ln\left[  
    \frac{2\overline{\K}\,\EXP{}}{\tilde{m}\,(-\lambda)\,(1-\theta_*)}
  \right]
\end{align}
which corresponds to 
\begin{align}
  \label{eq:MeanAknStandardCaseLargeK}
  \symmetriz{ a_{n+1}^{(k)} }_{\pi,\sigma}
  &\sim 
  \EXP{n\, \mean{\ln( m /\K)}}
  \left(  \frac{2n\EXP{}\overline{\K}}{\tilde{m}\,k'}\right)^{k'} 
   \mbox{ for }
   1\ll k'\ll n  
\end{align}
where $k'=n-k$.
Introducing the expression of $\theta_*$ in \eqref{eq:ProvExpForOmega}, we deduce the general behaviour  
%
\begin{align}
  \label{eq:Omega-LargeLambda}
  \Omega(\lambda) \simeq  
   \ln (-\lambda) + \mean{\ln(m/\K)}
    - \frac{2\overline{\K}}{ \tilde{m}\,\lambda}
  \mbox{ for }
  \lambda\to-\infty
  \:.
\end{align}
As for $\lambda\to0^-$, we could also expand the fugacity \eqref{eq:ExpansionFugacityLarge} to higher order in $(1-\theta)$. 
Clearly, this would give a systematic expansion of $\Omega(\lambda)$ involving integrer powers of $1/\lambda$.
Contrary to the $\lambda\to0^-$ expansion, the $\lambda\to-\infty$ is thus analytic in $1/\lambda$ (apart the leading logarithmic term).
Terms $\mathcal{O}(1/\lambda^n)$ do not contribute to the density of states~: in this case the formula \eqref{eq:MainResultForOmegaBis} only predicts the trivial asymptotic result $\IDoS(\lambda\to\infty)=1$. 
However when $\K_n<1$ and $m_n>1$, we expect a band edge with a rapid (exponential) fall off of the DoS (Lifshitz tail) \cite{LucNie88}~: see Fig~\ref{fig:IDoS}. 
Our formula \eqref{eq:MainResultForOmegaBis} seems unable to give information on such non analytic spectral singularity at the upper edge of the spectrum.

Cases where the main assumptions of this paragraph on $\mean{m_i}$, $\mean{\K_i}$, $\mean{1/m_i}$ and $\mean{1/\K_i}$ are not fulfilled will be discussed below.

\subsection{Test of the combinatorial method~: the case of exponentially distributed $m_j$'s and $1/\K_i$'s}

Let us first benchmark the method and consider a solvable case, when both the masses and the inverse spring constants are exponentially distributed~: $p(\K)=(a/K^2)\,\EXP{-a/\K}$ and $q(m)=b\,\EXP{-bm}$.
For this model, the complex Lyapunov exponent is known to be given by a ratio of MacDonald functions~(cf. Eq.~4.17 of Ref.~\cite{Nie84} or \cite{ComTexTou19})
\begin{equation}
  \label{eq:ComplexLyapForExpDistributions}
  \Omega(\lambda) = 
  \epsilon\, \frac{ K_0\big(2/\epsilon\big) }{ K_1\big(2/\epsilon\big) }
  \hspace{0.5cm}\mbox{with }
  \epsilon=\sqrt{\frac{-\lambda}{ab}}
  \:,
\end{equation}
where $\Kb=a$ and $\mb=1/b$.
Let us compare this exact formula with the approximate expression \eqref{eq:MainResultForOmega}, which we denote $\Omega_\mathrm{cm}(\lambda)$ in this paragraph.
As noticed above, for this model, the equation for the fugacity $\varphi_*^{(1/\K)}(\theta)$ has a simple form \eqref{eq:FugacityForExpVar}.
The second fugacity $\varphi_*^{(m)}(\theta)$ obeys the same equation for $a\to b$. 
As a result we can write $\varphi_*^{(1/\K)}(\theta)/a=\varphi_*^{(m)}(\theta)/b=\psi(\theta)$ where the function solves
\begin{equation}
  \label{eq:ExpCase-EqForPsi}
  \frac{1}{\psi(\theta)}\,\EXP{1/\psi(\theta)}\, E_1(1/\psi(\theta)) = 1-\theta
  \:.
\end{equation}
The saddle equation \eqref{eq:ThetaStar} is now
\begin{equation}
\label{eq:ExpCase-SaddleEq}
  \psi(\theta ) = \frac{\epsilon}{2}\sqrt{\frac{1+\theta}{1-\theta}}   
\end{equation} 
with solution denoted $\theta_*(\lambda)$.
We can use \eqref{eq:FreeEnergy} in this case~:
$\mean{\ln(1+\varphi/K_j)}=\EXP{a/\varphi} E_1(a/\varphi)=\EXP{1/\psi} E_1(1/\psi)$, so that \eqref{eq:MainResultForOmega} is
\begin{align}
  &\Omega_\mathrm{cm}= 
  (1+\theta_*)\ln(1+\theta_*)+(1-\theta_*)\ln(1-\theta_*)
  \\\nonumber
  &
  +2\theta_*\,\ln(\epsilon/2) 
  + 2 \EXP{1/\psi(\theta_*)} E_1(1/\psi(\theta_*)) - 2\theta_*\ln\psi(\theta_*)
  \:.
\end{align}
Combining this equation with (\ref{eq:ExpCase-EqForPsi},\ref{eq:ExpCase-SaddleEq}), we finally get the simple form
\begin{equation}
  \label{eq:ExpCase-Omega}
  \Omega_\mathrm{cm}(\lambda) 
  = \ln\left[ 1 - \theta_*(\lambda)^2 \right] + \epsilon\sqrt{1 - \theta_*(\lambda)^2}
  \:.
\end{equation}

Let us first study the $\lambda\to0^-$ limit.
We use a perturbative method~: 
we insert \eqref{eq:ExpCase-SaddleEq} in \eqref{eq:ExpCase-EqForPsi} and expand this equation in powers of $\epsilon$ and $\theta$. Then we look for the solution under the form $\theta=\theta_1+\theta_2+\cdots$, with $\theta_n=\mathcal{O}(\epsilon^n)$.
This allows to get $\theta_*(\lambda)$ order by order. 
Using the \texttt{Mathematica} software, we get the first terms of the expansion 
\begin{equation}
  \theta_*(\lambda) = 
   \frac12\epsilon
  -\frac14\epsilon^2
  +\frac{3}{16}\epsilon^3
  -\frac{5}{16}\epsilon^4
  +\frac{215}{256}\epsilon^5
  -\frac{175}{64}\epsilon^6
  +\mathcal{O}(\epsilon^7)
\end{equation}
Insertion in \eqref{eq:ExpCase-Omega} gives
\begin{equation}
  \Omega_\mathrm{cm}(\lambda) 
  =  \epsilon
  -\frac14\epsilon^2
  +\frac{1}{8}\epsilon^3
  -\frac{5}{32}\epsilon^4
  +\frac{43}{128}\epsilon^5
  -\frac{175}{192}\epsilon^6
  +\mathcal{O}(\epsilon^7)
\end{equation}
We compare this with the expansion of the exact result~\eqref{eq:ComplexLyapForExpDistributions} 
\begin{equation}
  \Omega(\lambda) 
  =  \epsilon
  -\frac14\epsilon^2
  +\frac{3}{32}\epsilon^3
  -\frac{3}{64}\epsilon^4
  +\frac{63}{2048}\epsilon^5
  -\frac{27}{1024}\epsilon^6
  +\mathcal{O}(\epsilon^7)
\end{equation}
The leading terms (dominant term of the IDoS) and next leading terms (dominant term of the Lyapunov exponent) exactly match.
However some discrepancy appears at third order, i.e. $\mathcal{O}((-\lambda)^{3/2})$.

We now study the limit $\lambda\to-\infty$.  
%
The dominant behaviour of the fugacity is $\psi(\theta)\simeq  (1-\theta)^{-1}\ln\big[\EXP{-\mathbf{C}}/(1-\theta)\big]$ for $\theta\to1$.
Due to the presence of logarithmic corrections, the analysis of the saddle equation \eqref{eq:ExpCase-SaddleEq} and the use of \eqref{eq:MainResultForOmegaBis}  beyond the leading contribution become tedious.
The detail of the calculation can be found in Appendix~\ref{app:ExpCase-ExpansionOmega-Large}.
Using $\mean{\ln(m/\K)}=-\ln ab-2\mathbf{C}$ and setting $\Lambda=(-\lambda)/ab$, we find 
\begin{align}
  \label{eq:OmegaForExpWeights}
&  \Omega_\mathrm{cm}(\lambda)
  \simeq
  \ln (\Lambda\,\EXP{-2\mathbf{C}})  + \frac{2}{\Lambda}
  \bigg\{
   \ln^2(\Lambda\,\EXP{-\mathbf{C}}/2)
  \\\nonumber
   &   
    \hspace{0.25cm} 
   +2\ln(\Lambda\,\EXP{-\mathbf{C}}/2) \, 
   \left[ 1 - \ln\ln(\Lambda\,\EXP{-\mathbf{C}}/2) \right] 
      + \mathcal{O}\!\left( (\ln\ln)^2 \right)
   \bigg\}   
\end{align}
for $\Lambda\to\infty$ (i.e. $\lambda\to-\infty$).

\begin{figure}[!h]
\centering
\includegraphics[width=0.4\textwidth]{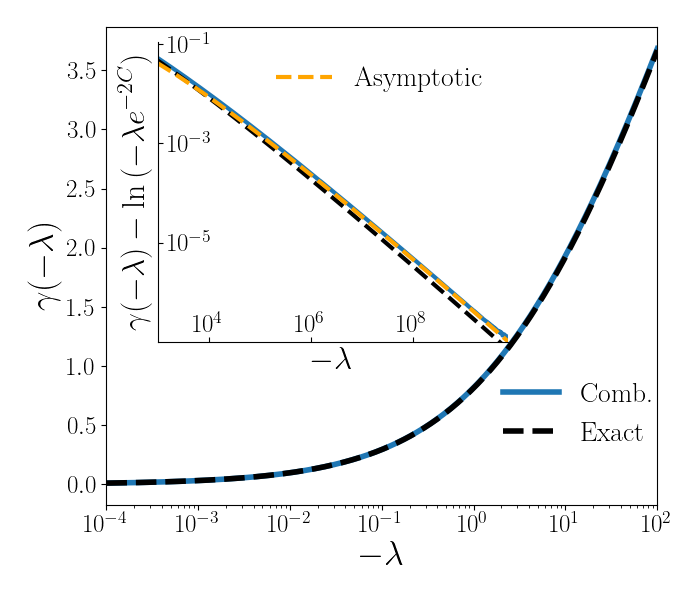}
\caption{\it Comparison between the exact result \eqref{eq:ComplexLyapForExpDistributions} (dashed line) and the approximate result of the combinatorial approach \eqref{eq:ExpCase-Omega} (continuous line), from a numerical resolution of the transcendental equation for $\theta_*$, i.e. Eqs.~(\ref{eq:ExpCase-EqForPsi},\ref{eq:ExpCase-SaddleEq}). 
Inset~: same with the dominant $\mathcal{O}(\ln(-\lambda))$ term substracted, with comparison to the asymptotics~\eqref{eq:OmegaForExpWeights} (orange dashed line).}
\label{fig:NumMax}
\end{figure}

We can compare this result with the expansion of the MacDonald functions in \eqref{eq:ComplexLyapForExpDistributions}, which gives
\begin{align}
  \label{eq:ExpCaseExactAsymp}
 &  \Omega(\lambda)
  \simeq
  \ln\left( \Lambda\,\EXP{-2\mathbf{C}} \right)  
 + \frac{1}{\Lambda}\,
 \bigg\{
   \ln^2\left( \Lambda\,\EXP{-2\mathbf{C}} \right)
 \nonumber\\
 & 
     \hspace{0.5cm} 
       +  2\,\ln\left( \Lambda\,\EXP{-2\mathbf{C}} \right)
       + 2
 +\mathcal{O}\left(\frac{\ln^2(\Lambda\,\EXP{-2\mathbf{C}})}{\Lambda}\right)
 \bigg\}
\end{align}
We have recovered the leading term, however the subleading corrections exhibit some discrepancies~; nevertheless the plot show that the two expressions remain very close, cf. Fig.~\ref{fig:NumMax}.

Note that after analytic continuation, the approximate $\Omega(\lambda+\I0^+)$ now exhibits some non trivial contribution to the spectral density which decays slowly (Fig.~\ref{fig:IDoS}). From \eqref{eq:ExpCaseExactAsymp}, 
\begin{equation}
  \label{eq:IDoS-exp-case}
  \IDoS(\lambda) 
  \underset{\lambda\to\infty}{\simeq}
  1 - \frac{2ab}{\lambda} \ln[\lambda\,\EXP{1-2\mathbf{C}}/ab]
 + \mathcal{O}\left(\frac{\ln\lambda}{\lambda^2}\right)
  \:.
\end{equation}
This slow decay of the spectral density, 
$\rho(\lambda)\sim
\ln(\lambda)/\lambda^2$ 
is due to the fact that here $\mean{\K}=\infty$ and $\smean{1/m}=\infty$ which induces high frequency modes.
This is very different from the rapid decay (Lifshitz tail) obtained for power-law distributions (\ref{eq:PowerLawK},\ref{eq:PowerLawMass}) with $\mean{\K}<\infty$ and $\smean{1/m}<\infty$~: see Fig~\ref{fig:IDoS}.
A careful study of Lifshitz tails for fixed spring constants and random masses $m_n>1$ was performed in \cite{LucNie88}.

%
%
%
%

We have tested the efficiency of the approach more precisely by solving 
(\ref{eq:ExpCase-EqForPsi},\ref{eq:ExpCase-SaddleEq})
numerically, injecting the solution $\theta_*(\lambda)$ in \eqref{eq:ExpCase-Omega}. 
In this way, we can determine the complex Lyapunov exponent in the full range of spectral parameter.
In Fig~\ref{fig:NumMax}, we compare the result of this procedure to the exact result \eqref{eq:ComplexLyapForExpDistributions}.
We see that the agreement is very good over several orders of magnitude.

In conclusion, despite the uncontrolled assumption on the symmetrization of the coefficients~$a_{n+1}^{(k)}$, the analysis of the case with exponential distributions shows that our combinatorial approach is able to give accurate results.


\section{Low frequency analysis}
\label{sec:LowFreq}

Low frequency properties of the spring chains seem the most interesting as we expect some universality in this regime.
The study of low energy properties of disordered systems has a long history~: the most famous result is probably the Lifshitz argument for the low energy spectral density of the Schr\"odinger equation \cite{Lif65} (see also the monographs \cite{LifGrePas88,Luc92}, or Kotani's paper \cite{Kot76} where a general analysis was given for repulsive potentials in 1D).
Concerning the localization properties, to the best of our knowledge, the first general result was demonstrated by Matsuda and Ishii \cite{MatIsh70} who considered spring chains with random masses and fixed spring constants $\K_n=\K$ $\forall\,n$.
These authors obtained a formula for the Lyapunov exponent at vanishing frequency
\begin{equation}
    \label{eq:MatsudaIshii1970}
    \gamma(\omega^2) \simeq \frac{\mean{\delta m_n^2}}{8\K\mean{m_n}}\omega^2
    \hspace{0.5cm} \mbox{for }\omega\to0
    \:,
\end{equation}
assuming $\mean{m_n^2}<\infty$.
The localization length thus grows as $\xi_\omega=1/\gamma(\omega^2)\propto\omega^{-2}$ at low frequency, i.e. the eigenmodes get delocalized as $\omega\to0$.
At the same time, the spectral density approaches the one for the perfect chain 
\begin{equation}
  \label{eq:DoSfiniteMeans}
  \rho(\lambda)\simeq \frac{1}{2\pi} \sqrt{\frac{\mean{m_n}}{\K\,\lambda}}
    \hspace{0.5cm} \mbox{for }\lambda\to0^+
    \:.
\end{equation}
These results come from the fact that mass disorder effectively vanishes in the $\omega\to0$ limit, as it is clear by inspection of Eq.~\eqref{eq:WaveEquation}.
Later, the random spring case (for $m_n=m$ $\forall\,n$) was considered in a different context \cite{BerSchWys80,AleBerSchOrb81,SteKar82} (1D master equation with random hopping rates).
The model considered by these authors corresponds to fixed masses, $m_n=1$ $\forall\,n$, and a power-law distribution for the spring constants
\begin{equation}
 \label{eq:PowerLawK}
 p(\K)=\mu\,\K^{-1+\mu}
 \hspace{0.5cm}\mbox{for }
  \K\in[0,1]
  \:,
\end{equation}
with $\mu>0$.
Note that we can always choose characteristic scales $\K\sim1$ and $m\sim1$ without loss of generality, as a rescaling of both $\K_n$'s and $m_n$'s can be reabsorbed in the frequency.
Bernasconi \textit{et al.} have observed a change in the low frequency spectral density from the pure case behaviour
$\rho(\lambda)\propto1/\sqrt{\lambda}$ for $\mu>1$, when $\mean{1/\K_n}=\mu/(\mu-1)<\infty$, to the power law
$\rho(\lambda)\propto\lambda^{\eta-1}$ for $\mu<1$, when $\mean{1/\K_n}=\infty$.
The exponent is $\eta=\mu/(1+\mu)\in\,]0,1/2]$, which shows that high probability of weak links (soft springs) leads to an accumulation of eigenmodes at small frequency.
When $\smean{1/\K_n}=\infty$, the effect of the disorder remains strong at small frequency, contrary to the case $\mean{1/\K_n}<\infty$, for which the disorder effectively vanishes as~$\omega\to0$.

Masses and inverse couplings play a symmetric role (cf. Section~\ref{sec:Duality}), hence, for the sake of generality, we will also consider a power-law distribution of masses
\begin{equation}
 \label{eq:PowerLawMass}
  q(m)=\nu\,m^{-1-\nu}
 \hspace{0.5cm}\mbox{for }
 m\in[1,\infty[
\end{equation}
with $\nu>0$.

For these distributions, we have $\mean{\K_i}<\infty$ and $\mean{1/m_i}<\infty$, thus the high frequency spectral and localization properties are encoded in \eqref{eq:Omega-LargeLambda} for $\mean{\K_i}=\mu/(\mu+1)$, $\mean{\ln\K}=-1/\mu$, $\mean{1/m_i}=\nu/(\nu+1)$ and $\mean{\ln m}=1/\nu$.
On the other hand, the low frequency regime is much richer~:
we will distinguish the cases 
$\mean{\K_n^{-1}}<\infty$ ($\mu>1$), $\mean{\K_n^{-1}}=\infty$ ($\mu\leq1$), $\mean{m_n}<\infty$ ($\nu>1$) and $\mean{m_n}=\infty$ ($\nu\leq1$), which will eventually allow us to draw a general picture on low frequency properties of disordered spring chains.

\subsection{Case $\mean{1/\K_n}<\infty$ and $\mean{m_n}<\infty$}

In this case, we can directly apply the generic result \eqref{eq:Omega-SmallLambda} obtained in the previous section.
We now use analytic continuation to get the spectral density.
When $\lambda<0$, the spectral density vanishes and the complex Lyapunov exponent coincides with the Lyapunov exponent, $\Omega(\lambda)=\gamma(\lambda)$, which characterizes the localization of the evanescent modes (which can exist only when the system has a boundary).
We can consider the case $\lambda>0$ by doing the substitution $\lambda\to\lambda+\I0^+$, which gives $\Omega(\lambda+\I0^+)\simeq -\I\sqrt{\lambda\mb/\Kb }$. The result is identified with $-\I\pi\,\IDoS(\lambda)$ according to \eqref{eq:OmegaAndDos}, hence 
\begin{equation}
  \label{eq:Dos-AlphaNegBetaNeg}
  \rho(\lambda) \simeq \frac{1}{2\pi}\sqrt{\frac{\mb}{\Kb \lambda}}
    \hspace{0.5cm}
  \mbox{for }
  \lambda\to0^+
  \:.
\end{equation}
For $\K_j=\K$ and small fluctuations of the masses, we recover~\eqref{eq:DoSfiniteMeans}.
This simple calculation, with analytic continuation has not provided the low energy Lyapunov exponent for $\lambda\to0^+$, Eq.~\eqref{eq:MatsudaIshii1970}, which requires the knowledge of the next leading order term of the $\lambda\to0^-$ expansion of $\Omega(\lambda)$. This will be further discussed below in Subsection~\ref{subsec:localization}.

\subsection{Case $\mean{1/\K_n}=\infty$ and $\mean{m_n}<\infty$}
\label{subsec:AlphaNegBetaPos}

We now analyze the case considered by Bernasconi \textit{et al.} \cite{BerSchWys80,AleBerSchOrb81,SteKar82}, where $\mean{1/\K_n}=\infty$ (exponent $\mu<1$), meaning that the probability for a soft spring (small $\K_j$) becomes large.

\subsubsection{A rough argument}

In order to understand in simple terms the origin of the power-law behaviour, let us first give a rough argument only relying on the expression of the coefficients \eqref{eq:MainRes2}.
We will provide a more precise analysis in a second step.
In \eqref{eq:MainRes2}, we can use \eqref{eq:AsympSymmProductSmallk} and simply replace the product of masses by $(\mb)^k$ since $\mean{m}<\infty$, leading to the form 
\begin{equation}
  a_{n+1}^{(k)} 
  \sim 
  (\mb)^k
  \hspace{-0.5cm}
  \sum_{1\leq i_1<\cdots<i_{k}\leq n}
   \ \prod_{m=1}^k 
   \frac{i_{m-1}-i_{m}}{\K_{i_{m}}} 
\end{equation}
We expect that the multiple sum is dominated by the contribution involving the set of $k$ smallest spring constants~:
\begin{equation}
  a_{n+1}^{(k)} 
  \sim 
  (\mb)^k \left( \frac nk \right)^k \big(\K_{i_1^*}\cdots\K_{i_k^*}\big)^{-1}
\end{equation}
and we have used that $(i_{m-1}-i_{m})\sim n/k$.
The largest inverse spring constant among the $n\gg1$ is typically $1/\K_{i_1^*}\sim n^{1/\mu}$, the second largest $1/\K_{i_2^*}\sim (n/2)^{1/\mu}$, etc. Hence 
\begin{equation}
  a_{n+1}^{(k)} 
  \sim 
  (\mb)^k \left( \frac nk \right)^k \left(\frac{n^{k}}{k!}\right)^{1/\mu}
  \sim\left( \frac{\mb}{\EXP{}}\right)^k \left( \frac{n\,\EXP{}}{k} \right)^{k/\eta}
\end{equation}
where 
\begin{equation}
  \eta = \frac{\mu}{\mu+1} 
  \ \in\,]0,1/2[
  \:.
\end{equation}
This coincides with the exponent of \cite{BerSchWys80,AleBerSchOrb81}.
The solution of \eqref{eq:SaddleForK} is now
\begin{equation}
  k_* \sim n \left(\frac{\mb(-\lambda)}{\EXP{}}\right)^\eta
\end{equation}
thus 
\begin{equation}
  x_{n+1}(\lambda)\sim a_{n+1}^{(k_*)}\,(-\lambda)^{k_*}\sim\EXP{k_*/\eta}
\end{equation}
 i.e. from \eqref{eq:LyapFromOptimum}, $\Omega\simeq k_*/(\eta\,n)$~:
\begin{equation}
  \Omega(\lambda) \sim  ( -\mb\lambda )^{\eta}
    \hspace{0.5cm}
  \mbox{for }
  \lambda\to0^-
  \:.
\end{equation}

\subsubsection{Detailed analysis}

\begin{figure}[!h]
\centering
\includegraphics[width=0.4\textwidth]{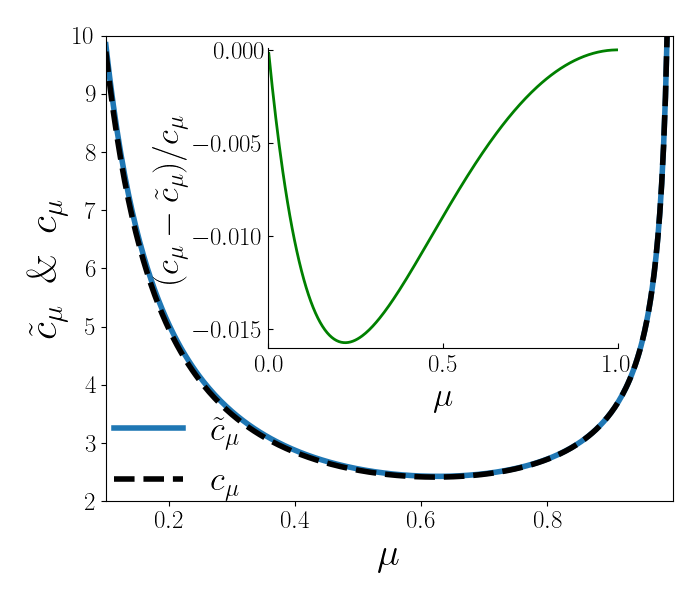}
\caption{\it 
  Coefficient $c_\mu$ (blue line) controlling the exact asymptotic behaviour and coefficient $\tilde{c}_\mu $ obtained from our combinatorial method (dashed red line). 
  Inset~:
  relative difference $(c_\mu-\tilde{c}_\mu)/c_\mu $ does not exceed 1.6\%.}
  \label{fig:CoeffCmu}
\end{figure}

To be more precise, we now apply \eqref{eq:MainResultForOmega}~:
from the expression of the fugacities for $\mu\in]0,1[$ and $\nu>1$, Eq.~\eqref{eq:FugacityForSmallTheta}, and \eqref{eq:ThetaStar}, we deduce 
\begin{equation}
  \theta_*(\lambda) \underset{\lambda\to0^-}{\simeq }
  \left( 
     \frac{\pi \mu  }{\sin\pi \mu£} 
  \right)^{1/(\mu+1)}
  \left( \frac{-\mb\,\lambda}{4}\right)^\eta 
\end{equation}
Inserting the approximate form of the fugacities in the integral \eqref{eq:MainResultForOmega} makes it extremely easy to compute 
\begin{align}
  \label{eq:112}
  \Omega\simeq 
  \theta_*\,\ln\left[ 
     \frac{\mb\,(-\lambda)}{4}\, \left( 
     \frac{\pi \mu  }{\sin\pi \mu} 
  \right)^{1/\mu}
  \left( \frac{\EXP{}}{\theta_*} \right)^{1/\eta}
  \right]
  \:,
\end{align}
corresponding to the asymptotics
\begin{equation}
  \symmetriz{ a_{n+1}^{(k)} }_{\pi,\sigma} \sim 
  \left( \frac{\mb}{4} \right)^k
 \left( 
     \frac{\pi \mu }{\sin\pi \mu}
  \right)^{k/\mu}
  \left(
      \frac{n\EXP{}}{k}
  \right)^{k(1+\mu)/\mu}
\end{equation}
for $k\ll n$.
Introducing the expression of $\theta_*$ in Eq.~\eqref{eq:112}, we conclude that $\Omega\simeq\theta_*/\eta$, i.e.
%
%
\begin{equation}
    \Omega(\lambda) \simeq  \tilde{c}_\mu\,( -\mb\,\lambda )^{\eta}
    \hspace{0.5cm}
  \mbox{for }
  \lambda\to0^-
\end{equation}
with 
\begin{equation}
  \tilde{c}_\mu =  \frac{4^{-\eta}}{\eta} \left( \frac{\pi \mu  }{\sin\pi \mu}   \right)^{1/(\mu+1)}
  \:.
\end{equation}
We can check the approximations by comparing this result with the \textit{exact} asymptotic obtained for a different model of disorder, when masses are exponentially distributed and with the same power-law distribution of spring constants~: in \cite{BerLeDRosTex24}, we obtained  $\Omega(\lambda) \simeq  c_\mu\,( -\mb\lambda )^{\eta}$ where
\begin{equation}
  c_\mu = 
  \frac{ \Gamma\big( \frac{\mu}{1+\mu} \big) \Gamma(1-\mu)^{1/(\mu+1)} }
       { (1+\mu)^{(1-\mu)/(1+\mu)}  \Gamma\big( \frac{1}{1+\mu} \big)  }
       \:.
\end{equation}
This coefficient was first obtained in \cite{SteKar82}.
Remarkably, the two coefficients $\tilde{c}_\mu $ and $c_\mu$ are very close, and differ at most by $1.6\,\%$ (Fig.~\ref{fig:CoeffCmu}).
The origin of this small discrepancy is in the replacement of the coefficient by the symmetrized one in the definition of the complex Lyapunov exponent, i.e. in the approximation $a^{(k)}_{n+1}\simeq\ssymmetriz{a^{(k)}_{n+1}}_{\pi,\sigma}$.

Using once again analytic continuation, we deduce the $\lambda\to0^+$ behaviour of the complex Lyapunov exponent.
Eventually 
\begin{equation}
  \label{eq:Lyap-AlphaPosBetaNeg}
  \gamma(\lambda) \simeq \cos(\pi\eta)\,\tilde{c}_\mu\,( \mb\lambda )^\eta
    \hspace{0.5cm}
  \mbox{for }
  \lambda\to0^+
\end{equation}
and 
\begin{equation}
  \label{eq:Dos-AlphaPosBetaNeg}
  \rho(\lambda) \simeq \frac{\eta\,\sin(\pi\eta)}{\pi}\,\tilde{c}_\mu\, \mb\, ( \mb\lambda)^{\eta-1}
    \hspace{0.5cm}
  \mbox{for }
  \lambda\to0^+
  \:.
\end{equation}
We have recovered the power law $\rho(\lambda) \sim  \lambda^{\eta-1} = \lambda^{-1/(1+\mu)}$ first obtained in \cite{BerSchWys80,SteKar82} (see also \cite{BerLeDRosTex24} where the power law was obtained by a more precise calculation for power-law distribution of spring constants and exponentially distributed masses).
The exponent of the Lyapunov exponent was first obtained in \cite{Zim82} for a related model (Eq.~24 of that paper).
Because $2\eta<1$, the density of modes $\varrho(\omega)=\D\IDoS(\omega^2)/\D\omega\sim \omega^{2\eta-1}$ is now divergent for $\omega\to0$ instead of constant,
i.e. there is an accumulation of low frequency modes due to power-law disorder.
Simultanously, delocalization of eigenmodes for $\omega\to0$ is slower than in the pure case~:~$\xi_\omega=1/\gamma(\omega^2)\propto\omega^{-2\eta}$.

\subsection{Marginal case $\mu=1$ (with $\mean{m_n}<\infty$)}

Eq.~\eqref{eq:ThetaStar} takes the form
\begin{equation}
   \label{eq:117}
  \frac{\theta_*^2}{\mb\,\ln(1/\theta_*)} \simeq \frac{(-\lambda)}{4} 
\end{equation}
 for $\theta\to0$,
i.e.
\begin{equation}
  \theta_*(\lambda) \simeq \frac12 \sqrt{-\mb\,\lambda \, \ln(2/\sqrt{-\mb\,\lambda})}
  \:.
\end{equation}
The integral \eqref{eq:MainResultForOmega} gives
\begin{align}
  \Omega \simeq 
  \theta_*\,\ln\left[
    \frac{\mb\,(-\lambda)}{4} \left(\frac{\EXP{}}{\theta_*}\right)^2 \ln(1/\theta_*)
  \right]
   + \mathrm{li}(\theta_*)
\end{align}
where $\mathrm{li}(x)=\int_0^x\D t/\ln t = -E_1(-\ln x)$ is the logarithmic integral.
We can neglect its contribution $\mathrm{li}(\theta_*)\simeq \theta_*/\ln \theta_*$, hence, making use of \eqref{eq:117} we obtain $\Omega\simeq2\theta_*$, i.e.
\begin{equation}
  \label{eq:Omega-AlphaZeroBetaNeg}
  \Omega(\lambda) \simeq \sqrt{-\mb\lambda\ln(2/\sqrt{-\mb\lambda})}
    \hspace{0.5cm}
  \mbox{for }
  \lambda\to0^-
  \:.
\end{equation}
Analytic continuation gives
\begin{equation}
  \label{eq:Lyap-AlphaZeroBetaNeg}
  \gamma(\lambda) \simeq \frac{\pi}{2}\sqrt{\frac{\mb\lambda}{\ln(4/\mb\lambda)}}
   \hspace{0.25cm} \mbox{for }
  \lambda\to0^+
\end{equation}
and
\begin{equation}
  \label{eq:Dos-AlphaZeroBetaNeg}
  \rho(\lambda) \simeq    \frac{1}{2\pi}\sqrt{\frac{\mb\ln(4/\mb\lambda)}{2\lambda}}
  \hspace{0.25cm}  \mbox{for }
  \lambda\to0^+
\end{equation}
Note that, up to the logarithmic correction, the exponent of $\gamma(\lambda)\sim\lambda^{1/2}$ matches with the one obtained for 
$\mu\in]0,1[$, $\gamma(\lambda) \sim\lambda^{\mu/(1+\mu)}$. 
Correspondingly, the density of modes is $ \varrho(\omega) \simeq\pi^{-1}\sqrt{\mb\,\ln(2/\sqrt{\mb}\,\omega)}$ for $\omega\to0$.

\subsection{Case $\mean{1/\K_n}<\infty$ and $\mean{m_n}=\infty$}
\label{subsec:AlphaPosBetaNeg}

This case $\mu>1$ and $\nu<1$ corresponds to a situation where the probability for very heavy mass becomes significant.
This case is simply symmetric to the case $\mu<1$ and $\nu>1$ studied above.
Using the symmetry argument (Section~\ref{sec:Duality}), we can replace $\mu\to\nu$ and $\mb\to1/\Kb$ in the results of \S~\ref{subsec:AlphaNegBetaPos}~:
we introduce now the exponent
\begin{equation}
  \eta = \frac{\nu}{1+\nu}
  \:.
\end{equation}
The complex Lyapunov exponent presents the behaviour 
$\Omega(\lambda) \simeq \tilde{c}_\nu\, ( -\lambda/\Kb )^{\eta}$ for $\lambda\to0^-$.  
Results (\ref{eq:Lyap-AlphaPosBetaNeg},\ref{eq:Dos-AlphaPosBetaNeg}) can be used upon substitution $\mu\to\nu$ and $\mb\to1/\Kb$, with now $\eta=\nu/(\nu+1)$.

\subsection{Marginal case $\nu=1$ (with $\mean{1/\K_n}<\infty$)}

We can as well perform 
$\mb\to1/\Kb$ in Eq.~\eqref{eq:Omega-AlphaZeroBetaNeg} and (\ref{eq:Lyap-AlphaZeroBetaNeg},\ref{eq:Dos-AlphaZeroBetaNeg}).

\subsection{Marginal case $\nu=1$ with $\mu<1$ ($\mean{1/\K_n}=\infty$)}

Eq.~\eqref{eq:ThetaStar} gives 
\begin{equation}
  \theta_* \simeq \left[ d_\mu\, (-\lambda)\, \ln1/\theta_* \right]^{\eta}
\end{equation}
where $\eta=\mu/(\mu+1)$ and $d_\mu=(\pi\mu/\sin\pi\mu)^{1/\mu}/4$.
The integral \eqref{eq:MainResultForOmega} gives
\begin{equation}
  \Omega\simeq \theta_*\,
  \left\{
    \ln \left[
      d_\mu\, (-\lambda)\,\left(\frac{\EXP{}}{\theta}\right)^{1/\eta} \ln1/\theta_*
    \right]
    +\mathcal{O}(1/\ln(1/\theta_*))
  \right\}
\end{equation}
leading again to $\Omega\simeq \theta_*/\eta$, i.e
\begin{equation}
  \Omega(\lambda) \simeq 
  \eta^{\eta-1}   \left( -d_\mu\,\lambda \, \ln[1/(-\lambda\,d_\mu)] \right)^\eta
\end{equation}
for $\lambda\to0^-$.
Analytic continuation then gives
$\gamma(\lambda)\sim\big[\lambda\,\ln(1/\lambda)\big]^\eta$ and $\rho(\lambda)\sim\lambda^{\eta-1}\,\big[\ln(1/\lambda)\big]^\eta$ for $\lambda\to0^+$.

\subsection{Marginal case $\mu=1$ with $\nu<1$ ($\mean{m_n}=\infty$)}

By symmetry we have 
\begin{equation}
  \Omega(\lambda) \simeq 
  \eta^{\eta-1}   \left( -d_\nu\,\lambda \, \ln[1/(-\lambda\,d_\nu)] \right)^\eta
  \:.
\end{equation}
where $\eta=\nu/(\nu+1)$.

\subsection{Doubly marginal case $\mu=1$ and $\nu=1$}

Clearly, the $\theta\to0$ limit of Eq.~\eqref{eq:ThetaStar} is now 
\begin{equation}
  \frac{\theta_*}{\ln(1/\theta_*)} \simeq \frac{\sqrt{-\lambda}}{2} 
\end{equation}
i.e.
\begin{equation}
   \label{eq:ThetaStarDoublyMarginal}
  \theta_*(\lambda) \simeq \frac14 \sqrt{-\lambda} \, \ln(4/(-\lambda))
  \:.
\end{equation}
The integral \eqref{eq:MainResultForOmega} gives
\begin{align}
  \Omega \simeq 
  \theta_*\,\ln\left[
    \frac{(-\lambda)}{4} \left(\frac{\EXP{}}{\theta_*}\ln(1/\theta_*)\right)^2 
  \right]
   + 2\,\mathrm{li}(\theta_*)
   \:.
\end{align}
Neglecting the contribution of the logarithmic integral and making use of \eqref{eq:ThetaStarDoublyMarginal} we get $\Omega\simeq2\theta_*$, i.e.
\begin{equation}
  \Omega(\lambda) \simeq \sqrt{ -\lambda }\, \ln (2/\sqrt{ -\lambda })
    \hspace{0.5cm}
  \mbox{for }
  \lambda\to0^-
  \:.
\end{equation}
Compared to \eqref{eq:Omega-AlphaZeroBetaNeg}, we see that the complex Lyapunov exponent has now received two logarithmic corrections. 
Analytic continuation gives this time
\begin{equation}
  \gamma(\lambda) \simeq \frac\pi2\,\sqrt{\lambda}
    \hspace{0.25cm}  \mbox{and}   \hspace{0.25cm}
  \rho(\lambda) \simeq \frac{\ln(4/\lambda)}{4\pi\sqrt{\lambda}}
    \hspace{0.5cm}
  \mbox{for }
  \lambda\to0^+
\end{equation}
i.e. the density of modes is $ \varrho(\omega) \simeq\pi^{-1} \ln(2/\omega)$ for $\omega\to0$.

\subsection{Case $\mean{1/\K_n}=\infty$ and $\mean{m_n}=\infty$}

In the case $\mu<1$ and $\nu<1$, Eq.~\eqref{eq:ThetaStar} gives 
\begin{equation}
  \theta_*(\lambda) \simeq \left[ A \,(-\lambda) \right]^\eta
\end{equation}
with 
\begin{equation}
  \eta = \left(  \frac{1}{\mu} + \frac{1}{\nu}  \right)^{-1}
\end{equation}
and 
\begin{equation}
  A = \frac{1}{4}
  \left( \frac{\pi\mu}{\sin \pi\mu} \right)^{1/\mu} 
  \left(\frac{\pi\nu}{\sin \pi\nu} \right)^{1/\nu}  
  \:.
\end{equation}
The exponent $\eta$ is continuous at $\mu=1$ and/or $\nu=1$~:
for example $\eta=\mu/(1+\mu)$ when $\nu=1$ and we recover $\eta=1/2$ when $\mu=\nu=1$.

Once more, the integral \eqref{eq:MainResultForOmega} can be straightforwardly calculated
\begin{equation}
  \Omega\simeq \frac{\theta_*}{\eta}\, \ln
  \left[
     \left[ A \,(-\lambda) \right]^\eta\,\frac{\EXP{}}{\theta_*}
  \right]
  = \frac{\theta_*}{\eta}
\end{equation}
i.e.
%
\begin{equation}
  \label{eq:Omega-AlphaPosBetaPos}
  \Omega(\lambda) \simeq C_{\mu,\nu}\, ( -\lambda )^{\eta}
    \hspace{0.5cm}
  \mbox{for }
  \lambda\to0^-
  \:.
\end{equation}
where 
\begin{equation}
  C_{\mu,\nu} = \frac{4^{-\eta}}{\eta}
    \left( \frac{\pi\mu}{\sin \pi\mu } \right)^{\nu/(\mu+\nu)} 
  \left(\frac{\pi\nu}{\sin \pi\nu } \right)^{\mu/(\mu+\nu)}  
\end{equation}
Analytic continuation gives
\begin{align}
  \gamma(\lambda) &\simeq \cos(\pi\eta)\,C_{\mu,\nu}\,\lambda ^\eta
    \hspace{1.15cm}
  \mbox{for }
  \lambda\to0^+
  \\
  \rho(\lambda) &\simeq \frac{\eta\,\sin(\pi\eta)}{\pi}\,C_{\mu,\nu}\, \lambda^{\eta-1}
    \hspace{0.5cm}
  \mbox{for }
  \lambda\to0^+
  \:.
\end{align}

\subsection{Localization properties for $\mean{1/\K_n}<\infty$ ($\mu>1$) and $\mean{m_n}<\infty$ ($\nu>1$)}
\label{subsec:localization}

We have obtained above the generic behaviour \eqref{eq:Omega-SmallLambda} when $\mean{1/\K_n}<\infty$ and $\mean{m_n}<\infty$.
The analytic continuation to positive $\lambda$ gives the purely imaginary result 
$\Omega(\lambda+\I0^+) \simeq -\I\sqrt{\lambda\mb/\Kb }$, 
which does not give information on the Lyapunov exponent, hence we must improve the approximation in order to get insight about localization.

As we have pointed out already, our method can provide $\Omega$ under the form of a systematic expansion when $\lambda\to0^-$ or $\lambda\to-\infty$.
The calculation takes the form of a perturbative expansion, where either $\lambda$ or $1/\lambda$ is the small parameter. 
Thus it is more flexible than standard weak disorder expansions, which are usually based on the Dyson-Schmidt integral equation and assume that the disorder has finite moments \cite{DerGar84,ZanDer88,Luc92} (see also \cite{ComLucTexTou13}, \S3) and therefore  
are not suited to power-law disorder (in Ref.~\cite{BieTex08}, a concentration expansion for the 1D Schr\"odinger equation with random impurities was shown to be more appropriate to the case of power-law disorder~; concentration expansion is discussed in the book \cite{LifGrePas88}).
Here, we will see that the precise nature of the disorder distribution does not affect the principle of the method.

\subsubsection{Case $\mean{\K_n^{-2}}<\infty$ ($\mu>2$) and $\mean{m_n^2}<\infty$ ($\nu>2$)}

The $\theta\to0$ behaviour of the two fugactities is given by \eqref{eq:FugacityForSmallTheta}.
Previously, we have obtained \eqref{eq:ThetaStarNormalCaseSmallLambda} by keeping only the leading terms of the fugacities.
If we keep the next to leading terms, Eq.~\eqref{eq:ThetaStar} reads
\begin{align}
  &\frac{ \theta^2 }{ \mean{\K^{-1}} \mean{m} }
  \left( 1 + \frac{ \mean{\K^{-2}} }{ \mean{\K^{-1}}^2 } \theta +\cdots\right)
  \left( 1 + \frac{ \mean{m^2} }{ \mean{m}^2 } \theta +\cdots\right)
  \nonumber\\
  =& \frac{(-\lambda)}{4}\,\frac{1+\theta}{1-\theta}
  \:.
\end{align}
Solution can be found by a perturbative method~:
\begin{equation}
  \theta_*(\lambda) \simeq \frac12\,\epsilon
  - \frac{B}{8}\, \epsilon^2
  + \mathcal{O}(\epsilon^{3})
\end{equation}
where $\epsilon=\sqrt{-\mb\,\lambda/\Kb }$ and 
\begin{equation}
  B = \frac{\smean{\K^{-2}}}{\smean{\K^{-1}}^2} + \frac{\smean{m^{2}}}{\smean{m}^2} - 2
  \:.
\end{equation}
The integral \eqref{eq:MainResultForOmega} is easy to compute
\begin{equation}
  \Omega \simeq 2\theta_*\,\ln\left[\frac{\EXP{}\,\epsilon}{2\theta_*}\right] - \frac{B}{2} \,\theta_*^2  +\mathcal{O}(\theta_*^{3})
  \:.
\end{equation}
%
Eventually, we get  
\begin{equation}
  \Omega(\lambda) \simeq \sqrt{\frac{-\mb\,\lambda}{\Kb}} + \frac{B}{8} \frac{\mb\,\lambda}{\Kb} 
  \hspace{0.5cm}\mbox{for } \lambda\to0^-
  \:.
\end{equation}
Analytic continuation to $\lambda>0$ contains a real part 
\begin{equation}
  \label{eq:LyapPertub}
  \gamma(\lambda) 
  \underset{ \lambda\to0^+ }{ \simeq }
  \frac{\smean{m} \smean{\K^{-1}}}{8}
  \left( 
     \frac{\smean{\K^{-2}}}{\smean{\K^{-1}}^2} + \frac{\smean{m^{2}}}{\smean{m}^2} - 2 
  \right)
  \lambda
\end{equation}
The result exhibits the symmetry $m_n\leftrightarrow1/\K_n$.
The perturbative result \eqref{eq:LyapPertub} was obtained earlier in \cite{Fog21}, although this paper did not point out the symmetry emphasized here.
As expected, the Lyapunov exponent vanishes in the weak disorder limit, when both $\mathrm{var}(K^{-1})$ and $\mathrm{var}(m)$ vanish.
For fixed spring constants, this formula coincides exactly with the well-known  Matsuda-Ishii formula \eqref{eq:MatsudaIshii1970}.
For the model with power-law distributions considered in the paper, Eq.~\eqref{eq:LyapPertub} reads explicitly 
\begin{equation}
  \label{eq:147}
  \gamma(\lambda) 
  \underset{ \lambda\to0^+ }{ \simeq }
  \frac{\mu\nu}{8(\mu-1)(\nu-1)}
  \left[
     \frac{1}{\mu(\mu-2)} + \frac{1}{\nu(\nu-2)}
  \right]
  \lambda
\end{equation}

\subsubsection{Case $\mean{\K_n^{-2}}=\infty$ ($1<\mu<2$) and $\mean{m_n^2}<\infty$ ($\nu>2$)}

From \eqref{eq:FugacityForSmallTheta}, the ``inverse spring constant fugacity'' is now 
\begin{equation}
  \varphi_*^{(1/\K)}(\theta) \simeq \Kb\,\theta \left( 1 + a_\mu\,\theta^{\mu-1} +\cdots \right)
  \hspace{0.5cm}\mbox{for } \theta\to0
\end{equation}
where 
\begin{equation}
  a_\mu = - \frac{ \pi\mu }{ \sin\pi\mu } \left( \frac{\mu-1}{\mu} \right)^\mu  >0
  \:.  
\end{equation}
As a result the saddle equation \eqref{eq:ThetaStar} is 
\begin{equation}
  \theta_* \left( 1 + \frac{a_\mu}{2}\,\theta_*^{\mu-1} + \mathcal{O}(\theta_*) \right) = \frac{\epsilon}{2} ( 1 + \mathcal{O}(\theta_*) )
  \,
\end{equation}
where $\epsilon=\sqrt{-\mb\,\lambda/\Kb }$. 
The $\mathcal{O}(\theta_*)$ in the parentheses of the l.h.s. comes from the fluctuations of the masses and is negligible here.
We deduce
\begin{equation}
  \theta_*(\lambda) =  \frac{\epsilon}{2} - \frac{a_\mu}{2} \left( \frac{\epsilon}{2} \right)^\mu + \mathcal{O}(\epsilon^2)
  \:.
\end{equation}
The integral \eqref{eq:MainResultForOmega} also gives $\Omega$ under the form of an expansion 
\begin{align}
  \Omega &= 2\theta_*\,\ln\left[ \frac{\epsilon}{2} \,\frac{\EXP{}}{\theta_*} \right] - \frac{a_\mu}{\mu}\, \theta_*^\mu + \mathcal{O}(\theta_*^2)
  \\
  &\simeq 2\theta_* + a_\mu \left( 1 - \frac1\mu \right) \theta_*^\mu
\end{align}
Combining the two expansions, we eventually get 
\begin{equation}
  \label{eq:Omega-IntermediateMuAndNu}
  \Omega(\lambda)\simeq \sqrt{\frac{-\mb\,\lambda}{\Kb}} - \frac{a_\mu}{\mu\,2^\mu} \left(\frac{-\mb\,\lambda}{\Kb}\right)^{\mu/2}
  \hspace{0.5cm}\mbox{for } \lambda\to0^-
  \:.
\end{equation}
In this calculation, we have neglected the contribution from the fluctuations of the mass, which gives here a negligible contribution $\mathcal{O}(\lambda)$ to $\Omega(\lambda)$.
Analytic continuation to positive $\lambda$ gives
\begin{equation}
  \label{eq:155}
  \gamma(\lambda) 
  \underset{ \lambda\to0^+ }{ \simeq } 
  b_\mu \,\big( \mb\Kb\, \lambda \big)^{\mu/2}
  \hspace{0.25cm}
  \mbox{where }
  b_\mu = \frac{\pi}{2^{\mu+1}\,\sin(\pi\mu/2)}
\end{equation}
where we have used $a_\mu=- \pi\mu {\Kb}^\mu / \sin\pi\mu $. 
The exponent $\mu/2$ coincides with a result obtained earlier in Ref.~\cite{Zim82} for a related model (Eq.~24 of that paper).
Note that the IDoS also receives a correction~: 
\begin{align}
\label{eq:CorrectionToIDoS}
  \IDoS(\lambda) \simeq \frac{1}{\pi} \sqrt{\frac{\mb\,\lambda}{\Kb}}
   + \frac{ \big( \mb\,\Kb \,\lambda \big)^{\mu/2}  }{2^{\mu+1} \cos(\pi\mu/2)}
\end{align}
The density of modes behaves as 
\begin{equation}
  \label{eq:CorrectionToDoS-mu-larger-1}
  \varrho(\omega) \simeq \varrho(0)-\kappa_\mu\,\omega^{\mu-1}
  \hspace{0.25cm}
  \mbox{for }
  \kappa_\mu=\frac{-\mu\big( \mb\,\Kb \big)^{\mu/2}}{2^{\mu+1}\cos(\pi\mu/2)}>0
\end{equation}
with $\varrho(0)=\pi^{-1}\sqrt{\mb/\Kb}$. 
This announces the divergence $\varrho(0)=\infty$ obtained for $\mu\leq1$ (when $\Kb=0$).

Note that (\ref{eq:CorrectionToIDoS},\ref{eq:CorrectionToDoS-mu-larger-1}) hold in fact in the range $\mu\in]1,3[$ (when $\nu>3$).
When $\mu\in]2,3[$ we find $\varphi^{(1/\K)}_*(\theta)=\Kb\,\theta\,\big[1+\Kb^2\mean{\K^{-2}}\theta + a_\mu\theta^{\mu-1}+\mathcal{O}(\theta^2)\big]$, from which we deduce $\theta_*(\lambda)=\epsilon/2-B\epsilon^2/8-(a_\mu/2)(\epsilon/2)^{\mu}+\mathcal{O}(\epsilon^3)$ and then 
$\Omega(\lambda)\simeq\epsilon-B\epsilon^2/8-(a_\mu/\mu)(\epsilon/2)^{\mu}+\mathcal{O}(\epsilon^3)$.
When $2<\mu<3$, the leading term of $\gamma(\lambda)$ is \eqref{eq:LyapPertub} with a correction $\mathcal{O}(\lambda^{\mu/2})$, however the IDoS is \eqref{eq:CorrectionToIDoS}.

\subsubsection{Case $\mean{\K_n^{-2}}<\infty$ ($\mu>2$) and $\mean{m_n^2}=\infty$ ($1<\nu<2$)}

Using again the symmetry argument, i.e. doing $\mb\leftrightarrow1/\Kb$ and $\mu\leftrightarrow\nu$ in \eqref{eq:155}, leads to  
\begin{equation}
  \label{eq:LyapForMu>2andNu<2}
  \gamma(\lambda) \simeq b_\nu \,(  \lambda/\mb\Kb )^{\nu/2}
  \hspace{0.5cm}
  \mbox{for }
  \lambda\to0^+
  \:.
\end{equation}

Interestingly, in this case the Lyapunov exponent can also be obtained by using a different argument~:
for fixed spring constant and random masses, the wave equation takes the form $(-m_n\lambda/K)\,x_n=x_{n+1}-2x_n+x_{n-1}$. 
The limit $\lambda\to0$ probes large scale properties, hence we can study the continuum limit, leading to consider the differential equation 
$\big(-E+V(x)\big)\,\psi(x)=\psi''(x)$ where 
$x=n\,\epsilon$ (here, $\epsilon\to0$ is a lattice spacing),
$\psi(x)=x_n$, 
$E=\mb\,\lambda/(K\epsilon^2)$ and $V(x)=-\delta m_n\,\lambda/(K\epsilon^2)$ where $\delta m_n=m_n-\mb$ and $\mb=\mean{m_n}$.
As recalled above, we cannot use standard weak disorder expansion for power-law disorder. 
However we can use the formula obtained in \cite{BieTex08} from a concentration expansion, which gives 
\begin{equation}
  \label{eq:BienaimeTexier2008}
  \gamma(\lambda) \simeq \frac12 \mean{\ln\left[1 + \frac{\lambda\,\delta m_n^2}{4\mb\,K}\right]}
\end{equation}
and is appropriate to study the limit $\lambda\to0^+$.
For $\mean{m_n^2}<\infty$ ($\nu>2$), we can expand the log and we recover precisely \eqref{eq:MatsudaIshii1970}.

For $\mean{m_n^2}=\infty$ with $\mean{m_n}<\infty$ (i.e. $1<\nu<2$), we cannot expand the log, however we can compute the average in \eqref{eq:BienaimeTexier2008} with the power law \eqref{eq:PowerLawMass}.
We find 
\begin{equation}
  \gamma(\lambda)\simeq 
  \frac\nu2 \int_1^\infty\frac{\D m}{m^{\nu+1}}\ln\left[1 + \frac{\lambda\,(m-\mb)^2}{4\mb\,K}\right]
  \:.
\end{equation}
In the $\lambda\to0$ limit, the integral is dominated by $m\lesssim\sqrt{4K/\lambda}$, hence the behaviour $\gamma(\lambda)\sim\lambda^{\nu/2}$.
In order to extract the precise coefficient of the power law, we consider 
\begin{align}
  &\left(\frac{4\mb K}{\lambda}\right)^{\nu/2}\gamma(\lambda)
  \nonumber\\
  &\simeq\frac{\nu}{2}\int_{(1-\mb)\sqrt{\lambda/(4\mb K)}}^\infty \frac{\D u\,\ln(1+u^2)}{\big(u+\sqrt{\mb\lambda/(4K)}\big)^{\nu+1}}
\end{align}
The integral has a limit for $\lambda\to0$, given by
\begin{equation}
  \int_0^\infty  \frac{\D u\,\ln(1+u^2)}{u^{\nu+1}} = \frac{\pi}{\nu\sin(\pi\nu/2)}
\end{equation}
Performing the subtitution $K\to\Kb$, we recover \textit{exactly} the result \eqref{eq:LyapForMu>2andNu<2} obtained with the combinatorial approach 
for random spring constants.
This observation shows that in this case, our combinatorial method provides the exact asymptotic result for the Lyapunov exponent at small~$\lambda$ [i.e. exact leading order and next leading order terms of $\Omega(\lambda)$].

In the marginal case $\nu=2$, 
the exponent $\nu/2$ matches with the one obtained for $\nu>2$, Eq.~\eqref{eq:147}, however
the Lyapunov exponent receives an additional logarithmic contribution \cite{BieTex08}, 
$\gamma(\lambda)\simeq [\lambda/8\K]\,\ln(4\EXP{}\K/\lambda)$ [which explains the divergence of the coefficient $b_\nu$ in \eqref{eq:LyapForMu>2andNu<2} when $\nu\to2^-$].

\subsubsection{Case $1<\mu<2$ and $1<\nu<2$}

Due to the perturbative nature of the calculation when $\mu>1$ and $\nu>1$, 
it is clear that $\Omega$ receives two corrections of the type of \eqref{eq:Omega-IntermediateMuAndNu}~:
\begin{equation}
    \Omega(\lambda)\simeq 
    \sqrt{\frac{-\mb\,\lambda}{\Kb}} 
    - \frac{a_\mu}{\mu\,2^\mu} \left(\frac{-\mb\,\lambda}{\Kb}\right)^{\mu/2}
    - \frac{a_\nu}{\nu\,2^\nu} \left(\frac{-\mb\,\lambda}{\Kb}\right)^{\nu/2}
\end{equation}
for $\lambda\to0^-$.
After analytic continuation we get
\begin{align}
 \gamma(\lambda) 
  \simeq
  b_\mu \,( \mb\Kb\, \lambda)^{\mu/2} +  b_\nu \,(  \lambda/\mb\Kb )^{\nu/2}
  \hspace{0.5cm}
  \mbox{for }
  \lambda\to0^+
  \:.
\end{align}
The smallest exponent dominates, hence we conclude that in this regime
\begin{equation}
  \gamma(\lambda) \sim \lambda^\zeta
  \hspace{0.5cm}\mbox{with }
  \zeta = \frac12\min{\mu}{\nu}
  \:.
\end{equation}
Similarly the density of modes is 
\begin{equation}
  \label{eq:Intermediate-Corr-DoS}
  \varrho(\omega)\simeq \varrho(0) - B_\zeta\,\omega^{2\zeta-1}
  \hspace{0.5cm}
  \mbox{for }
  \omega\to0
\end{equation}
with  $B_\zeta=\kappa_\mu>0$ for $\mu<\nu$ and $B_\zeta=\kappa_\nu/(\mb\Kb)^\nu$ for $\nu<\mu$.

\subsubsection{Numerics}

\begin{figure}[!h]
\centering
\includegraphics[width=0.45\textwidth]{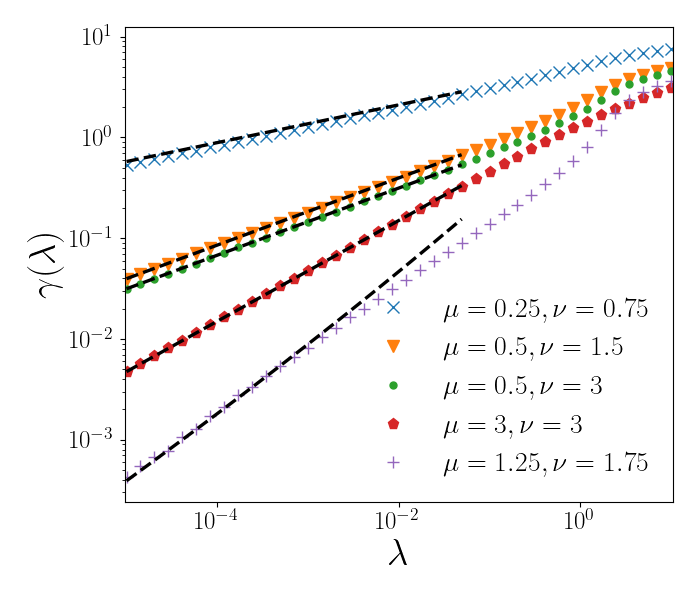}
\caption{\it 
  Lyapunov exponent obtained numerically for a chain of $10^5$ random masses and spring constants, as a function of $\lambda>0$ for different $\mu$ and $\nu$.
  Dashed lines correpond to the expected power law $\gamma(\lambda)\sim\lambda^\zeta$ (no adjustable parameter~: the prefactor is the one given in the text).  
  }
  \label{fig:LyapExponent}
\end{figure}

We have performed few numerical simulations for the Lyapunov exponent for different values of $\mu$ and $\nu$.
The results show a perfect agreement with the behaviour $\gamma(\lambda)\sim\lambda^\zeta$~:
cf. Fig.~\ref{fig:LyapExponent}.

\subsection{Discussion}

\begin{figure}[!h]
\centering
\includegraphics[width=0.35\textwidth]{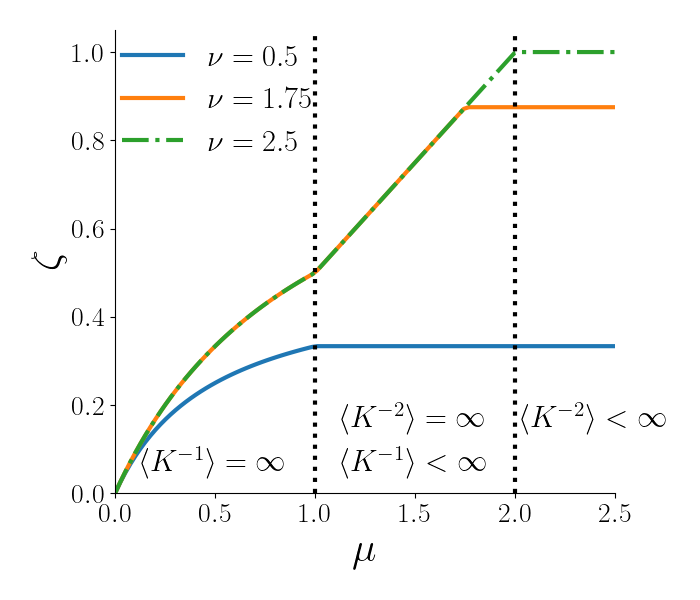}
\caption{\it Exponent $\zeta$ controlling the $\lambda\to0^+$ behaviour of the Lyapunov exponent as a function of $\mu$ for different values of~$\nu$.}
\label{fig:ZetaLyap}
\end{figure}

In this Section~\ref{sec:LowFreq}, we have studied in detail the low frequency properties of the random spring chains, which are the most interesting as they characterize large scale properties with expected universal character.
The spectral density and localization analysis allows to identify a "weak disorder regime", when $\mu>2$ and $\nu>2$ (finite $\mean{m^2}$ and $\mean{\K^{-2}}$) for which usual weak disorder expansions predicts at least the leading term~: 
the spectral density coincides with that of the perfect chain with $m\to\mean{m}$ and $\K\to1/\mean{\K^{-1}}$, explicitely
$\varrho(\omega)=\D\IDoS(\omega^2)/\D\omega\simeq\pi^{-1}\sqrt{\mean{m}\mean{\K^{-1}}}=\varrho(0)$ for $\omega\to0$.
The localization diverges at small frequency with the well-known behaviour $\xi_\omega\sim1/\omega^{-2}\to\infty$ \cite{MatIsh70}.
On the other hand, when $\mean{m^2}=\infty$ or $\mean{\K^{-2}}=\infty$, the disorder cannot be considered weak anymore (usual weak disorder expansions break down)~:
large fluctuations of the masses or the spring constants are responsible for anomalous localization $\xi_\omega\sim1/\omega^{-2\zeta}$ with exponent $\zeta<1$ (the eigenmodes exhibit localization on a scale which grows \textit{slower} than in the normal case as $\omega\to0$, indicating a stronger localization).
The exponent $\zeta$ was obtained for all possible power law disorder, which is summarized in Table~\ref{tab:Zeta} (Section~\ref{sec:Res}).
We can check that the exponent $\zeta$ is continuous, as shown in Fig.~\ref{fig:ZetaLyap} where it is plotted as a function of $\mu$ for different values of~$\nu$.

In the intermediate case where $\mean{m}<\infty$ and $\mean{\K^{-1}}<\infty$, with $\mean{m^2}=\infty$ or $\mean{\K^{-2}}=\infty$, the strong nature of the disorder only affects a subleading contribution of the spectral density~:
$\varrho(\omega)\simeq\varrho(0)-B_\zeta\,\omega^{2\zeta-1}$ with $B_\zeta>0$ and $2\zeta\in]1,2[$.

When fluctuations of masses or inverse spring constants are so large that all moments are infinite ($\mu<1$ and/or $\nu<1$), the density of modes diverges at small frequencies, $\varrho(\omega)\sim \omega^{2\eta-1}$ for exponent $2\eta\in]0,1[$ given in Table~\ref{tab:Eta}, and the localization length is $\xi_\omega\sim\omega^{-2\eta}$, which indicates a strong accumulation of localized eigenmodes at the band edge.


\section{Dyson's ``type I'': Anderson model with random couplings}
\label{sec:DysonTypeI}



In this last section, we discuss a case, known as "Dyson type I", which has attracted a lot of attention in the literature~:
when masses and spring constants exhibit some correlations in space leading to map the spring chain problem onto a model for a quantum particle on a lattice with random hoppings (Anderson model).
  Dyson  \cite{Dys53,For21} introduced the variables
  \begin{equation}
  \lambda_{2n-1} \eqdef \frac{\K_n}{m_n}
  \hspace{0.5cm}\mbox{and}\hspace{0.5cm}
  \lambda_{2n} \eqdef \frac{\K_n}{m_{n+1}}
  \end{equation}
and distinguished between ``type I'' and ``type II'' disorder.
Let us review and extend Dyson's classification~:
\begin{itemize}[leftmargin=*,align=left,itemsep=0.125cm]
\item
 Type I~: $\lambda_{j}$'s are i.i.d. random variables, which requires some unnatural correlations between $\K_n$'s and $m_n$'s.
 This model is the Anderson model for random couplings studied in this section. 
 It has been solved by Dyson when $\lambda_{j}$'s are distributed with a Gamma law~\cite{Dys53}.

\item
  Type II~: $\K_n=\K$ $\forall\,n$, while $m_n$'s are i.i.d. random variables. 
  Thus Dyson's parameters come in pairs as $\lambda_{2n}=\lambda_{2n+1}$.
  This case was studied by Schmidt \cite{Sch57} and many others (see Section~\ref{sec:Intro}).

\item
  Type II'~: $m_n=m$ $\forall\,n$ and $\K_n$'s are i.i.d. random variables.
  We have now  $\lambda_{2n-1}=\lambda_{2n}$
  (this case is dual to the type II).
  The spectral density was studied in \cite{BerSchWys80,AleBerSchOrb81} and the localization in~\cite{Zim82}.

\item
  Type III~: both $\K_n$'s and $m_n$'s are i.i.d. random variables. 
  This case was considered in Refs.~\cite{Nie84,BerLeDRosTex24} and in the present article 
  (note also a perturbative analysis for localization in Ref.~\cite{HerMen23}).
\end{itemize}

Starting with $N$ masses, Dyson's analysis for ``type I'' has relied on a mapping of the $N$ equations \eqref{eq:WaveEquation} onto the tight binding equation describing the Anderson model with random couplings~:
\begin{equation}
  \label{eq:AndersonRandomCouplings}
  - \rh_{n}^*\, \psi_{n+1}  - \rh_{n-1}\, \psi_{n-1}  = \omega\, \psi_{n}
  \:,
\end{equation}
with $n=1,\cdots,2N-1$. 
The couplings are related to the original parameters as $\rh_{n}=\I\sqrt{\lambda_n}$ \cite{Dys53}.
In the following we will rather consider Eq.~\eqref{eq:AndersonRandomCouplings} for real couplings $\rh_{n}$, as the phase plays no role.
This model has been much studied in the literature, from the perspective of quantum localization \cite{Bus75,TheCoh76,EggRie78,Dha80,Zim82}.

We now apply the combinatorial method to this case.
The spectral problem is defined by boundary conditions $\psi_0=\psi_{2N+1}=0$, which leads to $N$ pairs of eigenvalues $\pm\omega_\alpha$.
The fact that the solutions of the spectral problem \eqref{eq:AndersonRandomCouplings} come in pairs $\pm\omega_\alpha$ reflects the symmetry in (\ref{eq:WaveEquation},\ref{eq:AndersonRandomCouplings}) and explains the doubling of the number of equations~\cite{For21}.
We now study the initial value problem, i.e. we set $  \psi_0=0 $ and $\psi_1=1$ and solve the recurrence 
\begin{equation}
  \label{eq:RecurrenceAMOD}
  \psi_{n+1}(\omega)  = -\frac{\omega}{\rh_n}\, \psi_{n}(\omega) -\frac{ \rh_{n-1}}{\rh_n}\, \psi_{n-1}(\omega)
  \:.
\end{equation}
It is instructive to write the first terms (for $\lambda=-\omega$)~:
\begin{align}
  \psi_2(-\lambda) &= \frac{\lambda}{\rh_1}
  \\
  \psi_3(-\lambda) &= 
  \frac{-\rh_1}{\rh_2} + \frac{\lambda}{\rh_2}\,\frac{\lambda}{\rh_1}
  \\
  \psi_4(-\lambda) &= 
  \frac{\lambda}{\rh_3}\,\frac{-\rh_1}{\rh_2}
  +\frac{-\rh_2}{\rh_3}\,\frac{\lambda}{\rh_1}
  + \frac{\lambda}{\rh_3}\,\frac{\lambda}{\rh_2}\,\frac{\lambda}{\rh_1}
  \\
  \psi_5(-\lambda) &= 
   \frac{-\rh_3}{\rh_4}\,\frac{-\rh_1}{\rh_2}
   + \frac{\lambda}{\rh_4}\,\frac{\lambda}{\rh_3}\,\frac{-\rh_1}{\rh_2}
   + \frac{\lambda}{\rh_4}\,\frac{-\rh_2}{\rh_3}\,\frac{\lambda}{\rh_1}
   \nonumber\\
   &\hspace{1cm}
   + \frac{-\rh_3}{\rh_4}\,\frac{\lambda}{\rh_2}\,\frac{\lambda}{\rh_1}
   + \frac{\lambda}{\rh_4}\,\frac{\lambda}{\rh_3}\,\frac{\lambda}{\rh_2}\,\frac{\lambda}{\rh_1}
\end{align}
etc. 
For an even number of sites, writing \eqref{eq:AndersonRandomCouplings} as $\sum_mH_{nm}\psi_m=\omega\psi_n$ 
where $H$ is a $2N\times2N$ tridiagonal matrix, we have 
\begin{equation}
  \psi_{2N+1}(\omega) 
  = \frac{\prod_{\alpha=1}^N(\omega^2-\omega_\alpha^2)}{\prod_{j=1}^{2N}\rh_j} 
  = \frac{\det\left( \omega\,\mathbf{1}_{2N} - H \right)}{\prod_{j=1}^{2N}\rh_j}
  \:,
\end{equation}
which emphasizes once more that $\psi_{2N+1}(\omega)$ is also the spectral determinant of the tight binding Hamiltonian $H$
(see appendices of Refs.~\cite{GraTexTou14,FyoLeDRosTex18}).
For an odd number of sites, we have 
\begin{equation}
   \psi_{2N}(\omega) 
  = -\frac{\omega\prod_{\alpha=1}^{N-1}(\omega^2-\omega_\alpha^2)}{\prod_{j=1}^{2N-1}\rh_j} 
  = -\frac{\det\left( \omega\,\mathbf{1}_{2N-1} - H \right)}{\prod_{j=1}^{2N-1}\rh_j}
\end{equation}
This illustrates that for an even number of sites, eigenfrequencies comes in pairs $\pm\omega_\alpha$ with $\omega_\alpha>0$, while for an odd number of sites, there is an additional eigenvalue $\omega_0=0$ (the existence of a zero mode is a topological effect).

Given the solution, the aim is to analyze the complex Lyapunov exponent
\begin{equation}
  \Omega(\omega) = \lim_{N\to\infty} \frac{\ln \psi_{N+1}(\omega)}{N}
  \:.
\end{equation}

\begin{figure}[!h]
\centering
\includegraphics[scale=0.6]{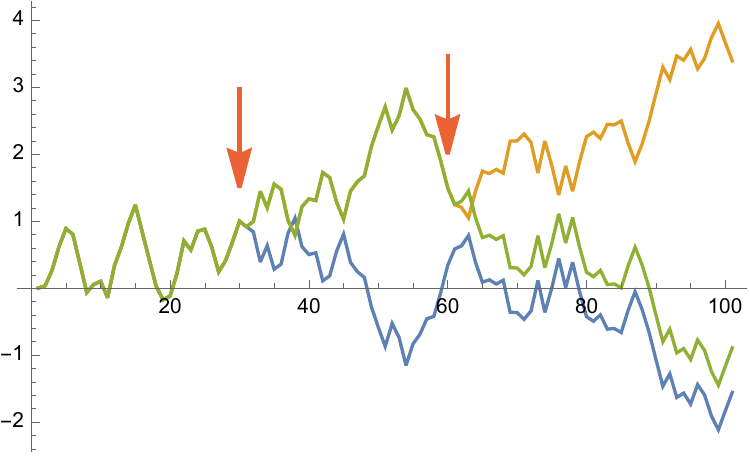}
\caption{\it 
The three random walks
$B_n$ (blue), $B_n(j_1)$ (orange) and $B_n(j_1,j_2)$ (green). 
The arrows indicate the positions of $j_1$ and $j_2$.
}
\label{fig:ExpB}
\end{figure}

As before, we write the solution of the initial value problem as a polynomial in~$\omega$~:
\begin{equation}
  \psi_{n+1}(\omega) = \sum_{k=0}^n \psi_{n+1}^{(k)}\,(-\omega)^k
\end{equation}
Inspection of the recurrence \eqref{eq:RecurrenceAMOD} (or the above determinantal representations) shows that 
$\psi_{2n+1}(-\omega)=\psi_{2n+1}(\omega)$ and $\psi_{2n}(-\omega)=-\psi_{2n}(\omega)$. 
As a consequence, the odd coefficients for even number of sites vanish $\psi_{2n+1}^{(2j+1)}=0$, and similarly $\psi_{2n}^{(2j)}=0$.
We have
\begin{widetext}
\begin{align}
  \psi_{2n+1}^{(0)} & = \frac{-\rh_{2n-1}}{\rh_{2n}}\cdots\frac{-\rh_3}{\rh_4}\,\frac{-\rh_1}{\rh_2}
  = \prod_{j=1}^n \frac{-\rh_{2j-1}}{\rh_{2j}}
  \hspace{0.5cm}\mbox{and } \psi_{2n}^{(0)}=0
  \\
    \psi_{2n}^{(1)} & 
    =\sum_{j=0}^{n-1} 
    \frac{-\rh_{2n-2}}{\rh_{2n-1}} \cdots\frac{-\rh_{2j+2}}{\rh_{2j+3}}\,
    \frac{1}{\rh_{2j+1}}\,
    \frac{-\rh_{2j-1}}{\rh_{2j}}\cdots\frac{-\rh_3}{\rh_4}\,\frac{-\rh_1}{\rh_2} 
  \hspace{0.5cm}\mbox{and } \psi_{2n+1}^{(1)}=0
  \\
    \psi_{2n+1}^{(2)} & 
    =\sum_{j_1>j_2}  
    \frac{-\rh_{2n-1}}{\rh_{2n}}\cdots
    \frac{-\rh_{2j_1+1}}{\rh_{2j_1+2}}\,
    \frac{1}{\rh_{2j_1}}\,
    \frac{-\rh_{2j_1-2}}{\rh_{2j_1-1}} \cdots\frac{-\rh_{2j_2+2}}{\rh_{2j_2+3}}\,
    \frac{1}{\rh_{2j_2+1}}\,
    \frac{-\rh_{2j_2-1}}{\rh_{2j_2}}\cdots\frac{-\rh_3}{\rh_4}\,\frac{-\rh_1}{\rh_2} 
  \hspace{0.5cm}\mbox{and } \psi_{2n}^{(2)}=0
\end{align}
\end{widetext}
etc.
It is interesting to introduce the notation 
\begin{equation}
  w_n \eqdef \ln \rh_n
\end{equation}
We can rewrite
\begin{align}
  \psi_{2n+1}^{(0)} & = (-1)^n \,\EXP{ B_n }
  \\
    \psi_{2n}^{(1)} & 
    =(-1)^{n-1} \sum_{j_1=0}^{n-1} \EXP{ B_n(j_1) }
      \\
    \psi_{2n+1}^{(2)} & 
    =(-1)^n \sum_{0\leq j_2<j_1\leq n-1}   \EXP{ B_n(j_1,j_2) }
\end{align}
etc, where 
\begin{align}
 \label{eq:Bn0}
 B_n &= \sum_{k=1}^{n} (-w_{2k} + w_{2k-1})
 \\
 B_n(j_1) &= 
 \sum_{k=j_1+1}^{n-1} (-w_{2k+1} + w_{2k})
 -w_{2j_1+1}
 \nonumber\\
 &
 +\sum_{k=1}^{j_1} (-w_{2k} + w_{2k-1})
 \label{eq:Bn1}
 \\
 B_n(j_1,j_2) &=
 \sum_{k=j_1+1}^{n} (-w_{2k} + w_{2k-1})
 - w_{2j_1}
 \nonumber\\
 &
  + \sum_{k=j_2+1}^{n-1} (-w_{2k+1} + w_{2k})
 -w_{2j_2+1}
 \nonumber\\
 &
 +\sum_{k=1}^{j_2} (-w_{2k} + w_{2k-1})
 \label{eq:Bn2}
\end{align}
etc.
Considering $B_n$ as a random walk, we see that each insertion of an index $j_m$ corresponds to introduce a reflection in the random walk (see Fig.~\ref{fig:ExpB}).

Being interested in an even number of sites, we should consider random walks with $2r$ reflections
\begin{equation}
   \psi_{2n+1}^{(2r)} 
    =(-1)^{n-r} \sum_{j_{2r}<\cdots< j_2<j_1}   \EXP{ B_n(j_1,j_2,\cdots,j_{2r}) }
\end{equation}
This picture makes easy to estimate the dominant contribution to
\begin{align}
  \label{eq:Psi2N}
  &\psi_{2n+1}(-\lambda) = \sum_{r=1}^n \psi_{2n+1}^{(2r)}\,\lambda^{2r}
  \\\nonumber
  = &\sum_{r=1}^n   (-1)^{n-r} \lambda^{2r}\sum_{j_{2r}<\cdots< j_2<j_1}   \EXP{ B_n(j_1,j_2,\cdots,j_{2r}) }
\end{align}
The term which dominates the sum is the one where the $2r$ reflections are such that the $2r$ pieces of $B_n(j_1,j_2,\cdots,j_{2r})$ are positive. 
For small $\lambda$, we expect that the sum is dominated by the terms with small~$r/n$.

\subsection{Case $\mean{(\ln \rh_j)^2} < \infty$}

We assume that
\begin{equation}
  \mean{(\ln \rh_j)^2} < \infty
\end{equation}
so that $B_n$ is a normal symmetric random walk, i.e. it scales as $B_n\sim\sqrt{n}$ and $\mean{B_n}=0$.
The $2r$ reflections split the random walk into $2r+1$ pieces, each piece being of order $\sim\sqrt{2n/2r}$.
Among the $2^{2r}$ trajectories obtained by introducing $2r$ reflections, there exists one optimal case for which the $2r+1$ pieces are all \textit{positive} (see Fig.~\ref{fig:ExpB})~:
\begin{equation}
  B_n(j_1,j_2,\cdots,j_{2r})\big|_{\mathrm{optimal}} \sim 2r \,\sqrt{n/r} = 2\sqrt{n\,r}
\end{equation}
Inserting this behaviour in the sum \eqref{eq:Psi2N}, we see that the sum is dominated by the optimal $r$ solution of 
\begin{equation}
  \deriv{}{r} \left(2r\,\ln(\lambda) + 2\sqrt{n\,r} \right) = 0
\end{equation}
so that $r_*\approx n/\big[4\ln^2(1/\lambda)\big]$, which is consistent with the assumption on $r$.
Correspondingly 
\begin{equation}
  \label{eq:Susy-Psi2n+1}
  \psi_{2n+1}(-\lambda)  \sim \EXP{2[r_*\,\ln(\lambda) + \sqrt{n\,r_*}]}
  \sim \exp\left\{n/\big[2\ln(1/\lambda)\big]\right\}
\end{equation}
Note that we have neglected in $B_n(j_1,\cdots,j_{2r})$ the contributions of the $2r$ ``isolated'' $-w_{2j'}$'s, cf. (\ref{eq:Bn1},\ref{eq:Bn2}). They contribute to $B_n(j_1,\cdots,j_{2r})$ typically as $-2r_*\mean{w_{j}}\sim n\mean{w_{j}}/\ln^2(\lambda)$, which is indeed negligible compared to $n/\ln(1/\lambda)$ in the exponential of Eq.~\eqref{eq:Susy-Psi2n+1}.
Finally, we deduce
\begin{equation}
  \label{eq:DysonSingNeg}
  \Omega(\omega)=\gamma(\omega)\sim \frac{1}{2\ln[1/(-\omega)]}
  \hspace{0.5cm}
  \mbox{for }
  \omega\to0^-
  \:.
\end{equation}
Analytic continuation to $\omega>0$ gives 
\begin{equation}
  \label{eq:DysonSing}
   \gamma(\omega) \sim \frac{1}{\ln(1/\omega)}
    \hspace{0.25cm}  \mbox{and}   \hspace{0.25cm}
  \IDoS(\omega) \sim \frac{1}{\ln^2(1/\omega)}
   \hspace{0.25cm} \mbox{for }
  \omega\to0^+
  \:. 
\end{equation}
We have recovered the expected Dyson singularity $\varrho(\omega)=\IDoS'(\omega)\sim1/\big[\omega\ln^3(1/\omega)\big]$ obtained in~\cite{Dys53,TheCoh76,EggRie78,Dha80,Zim82}.

\subsection{Case $\mean{(\ln \rh_j)^2} = \infty$}

The above results can be extended to the case where $\langle (\ln{t_j})^2 \rangle = \infty$. First, consider the scenario in which $\langle \ln{t_j} \rangle < \infty$. In this case, $\langle B_n \rangle = 0$ and $B_n \sim n^{1/\mu}$ with $1 < \mu < 2$. For the optimal choice of reflexions, one finds
$B_n(j_1, \dots, j_{2r})|_{\textrm{optimal}} \sim 2r \left({n}/{r}\right)^{1/\mu}$.
Applying the same method as before, the critical value $r_*$ is found to be $r_* = {n}/{\big[ (1 - 1/\mu) \ln{(1/\lambda)} \big]^\mu}$, which, for $\omega < 0$, yields 
\begin{equation}
\Omega(\omega)=\gamma(\omega) \sim \frac{1}{  \ln^\mu(1/(-\omega)) },
\end{equation}
extending Eq.~\eqref{eq:DysonSingNeg}. 
By analytic continuation, one obtains
\begin{equation}
\gamma(\omega) \sim \frac{1}{ \ln^\mu(1/\omega) }
\quad \textrm{and} \quad
\mathcal{N}(\omega) \sim \frac{1}{ \ln^{2\mu}(1/\omega)}.
\end{equation}
for $\omega\to0$.

Finally, in the case where $\langle \ln t_j \rangle = \infty$, the sum $B_n$ is dominated by its largest element, which scales as $n^{1/\mu}$ for $0 < \mu < 1$. The optimal choice of $r$ reflections is obtained by ensuring that the $2r$ largest elements contribute positively. Since the $j$th largest element scales as $(n/j)^{1/\mu}$, one finds that the optimal contribution is
\begin{equation}
B_n(j_1, \dots, j_{2r})\big|_{\textrm{optimal}} \sim \sum_{j=1}^r \left(\frac{n}{j}\right)^{1/\mu} \sim n^{1/\mu},
\end{equation}
as the sum converges for $\mu < 1$. Consequently, the maximum depends only weakly on $r$, and since $r \ln \lambda$ decreases with $r$, the optimal value is $r_* \sim 1$.
Thus, the leading behaviour of $\ln \psi_n(-\lambda)$ is
\begin{equation}
\ln \psi_n(-\lambda) \sim 2 \ln(\lambda) + (2n)^{1/\mu} \sim n^{1/\mu}.
\end{equation}
Such an anomalous scaling of the wavefunction $\ln\psi_n\sim n^{1/\mu}$ (instead of  $\ln\psi_n\sim n$) characterizes superlocalisation (see  for example \cite{BieTex08} and references therein).


\section{Conclusion}
\label{sec:Conclu}

We have analyzed the spectral and localization properties of random spring chains.
The interest in such discrete one-dimensional disordered systems goes back to the early steps of the theory of disordered wave equations, however the standard approach (Dyson-Schmidt method) requires the analysis of some integral equation which is in general extremely difficult. 
A perturbative analysis (weak disorder expansion) is possible 
\cite{DerGar84,ZanDer88,Luc92,HerMen23},
however it is not suited to study power-law random variables considered here.
For these reasons, the full phase diagram of spring chains with both random masses and random springs was not fully understood so far.

In the present article, we have proposed a new combinatorial approach based on the representation of the solution of the wave equation as a series of the spectral parameter, $x_{n+1}(\lambda)=\sum_{k=0}^n a_{n+1}^{(k)}\,(-\lambda)^k$, and an asymptotic analysis of the coefficients
\cite{footnote4}.  
This has been demonstrated to be efficient for different models (random spring chains, Anderson model with random couplings,...) and various disorder distributions. 
For the spring chain model, introducing an additional symmetrization with respect to the i.i.d. random parameters (masses and spring constants), we have obtained a new compact and general formula for the complex Lyapunov exponent
\eqref{eq:MainResultForOmega} [or \eqref{eq:MainResultForOmegaBis}].
It is interesting to note that thanks to the self averaging character of the complex Lyapunov exponent, the symmetrization of 
$\Omega(\lambda)=\lim_{N\to\infty}(1/N)\ssymmetriz{\ln\big(\sum_{k=0}^Na_{N+1}^{(k)}(-\lambda)^k\big)}_{\pi,\sigma}$ is exact.
Our approximation can be interpreted as considering the logarithm of the symmetrized sum, i.e.
$\Omega(\lambda)\simeq\lim_{N\to\infty}(1/N)\ln\big(\ssymmetriz{\sum_{k=0}^Na_{N+1}^{(k)}(-\lambda)^k}_{\pi,\sigma}\big)$ leading to  \eqref{eq:MainResultForOmega}.
Although our main result \eqref{eq:MainResultForOmega}
is not exact, it allows to get many accurate results rather easily~: 
the limiting behaviours of the complex Lyapnuov exponent both at small and large frequency can be obtained, involving relatively simple calculations.
While it is often sufficient to determine the leading term (at low or high frequency), we have shown that the method can provide  systematic expansions at low and high frequency.
At low frequency, 
we have obtained the limit of the eigenmode density
\begin{equation}
 \varrho(\omega)=2\omega\,\rho(\omega^2)
  \underset{ \omega\to0}{ \simeq }
 \frac{\sqrt{\mean{m}\mean{1/\K}}}{\pi} =\varrho(0)
\end{equation}
for $\mean{m}<\infty$ and $\mean{1/\K}<\infty$.
For $\mean{m^2}<\infty$ and $\mean{1/\K^2}<\infty$, the Lyapunov exponent decays at small frequency as
\begin{equation}
  \gamma(\omega^2)
    \underset{ \omega\to0}{ \simeq }
  \frac{\mean{m}\mean{1/\K}}{8}
  \left(
    \frac{\mathrm{var}(1/\K)}{\mean{1/\K}^2} + \frac{\mathrm{var}(m)}{\mean{m}^2} 
  \right)
  \omega^2
\end{equation}
while for large $\omega$, 
assuming $\mean{1/m}<\infty$ and $\mean{\K}<\infty$,
it behaves as
\begin{equation}
  \gamma(\omega^2) \underset{\omega\to\infty}{\simeq}  
  \mean{\ln(\omega^2\, m/\K)}  - \frac{2\smean{\K}\smean{m^{-1}}}{\omega^2}+ \mathcal{O}(1/\omega^{4})
  \:.
\end{equation}
In this case, the high frequency DoS is expected to decay fast and present Lifshitz tail, what the method is not able to predict.
The origin of this is most probably in the symmetrization of the coefficients $a_{n+1}^{(k)}$ with respect to the disorder parameters, which has led to our new formulae \eqref{eq:MainResultForOmega} and \eqref{eq:MainResultForOmegaBis}. 
The justification of this approximation certainly requires further investigation.
Still, we explain the absence of Lifshitz tail in our result by the following remark~: Lifshitz tails are due to atypical disorder configurations responsible for high frequency modes, thus the symmetrization of the coefficients prevents such configurations to have a prominent contribution.

These results show in particular that the disorder effectively vanishes at low frequency when $\mean{m}<\infty$ and $\mean{1/\K}<\infty$ as we recover the properties of the perfect spring chain. 
This is not the case when $\mean{m}=\infty$ ($\nu<1$) or $\mean{1/\K}=\infty$ ($\mu<1$), where disorder remains relevant as $\omega\to0$. 
We have obtained 
\begin{equation}
  \varrho(\omega) \sim \omega^{2\eta-1}
  \hspace{0.5cm}\mbox{and}\hspace{0.5cm}
  \gamma(\omega^2)\sim\omega^{2\eta}
\end{equation}
with the exponent $\eta$ given in Table~\ref{tab:Eta}.
Then the exponent is such that $\eta\in]0,1/2[$, which shows that the strong disorder induces an accumulation of low frequency eigenmodes ($\varrho(\omega)\to\infty$ for $\omega\to0$).
In the region of the phase diagram where $\mean{m}<\infty$ and $\mean{1/\K}<\infty$ with $\mean{m^2}=\infty$ or $\mean{\K^{-2}}=\infty$, we obtained 
\begin{equation}
\label{eq:206}
  \varrho(\omega) \simeq \varrho(0) -B_\zeta\, \omega^{2\zeta-1}
  \hspace{0.5cm}\mbox{and}\hspace{0.5cm}
  \gamma(\omega^2)\sim\omega^{2\zeta}
\end{equation}
with $B_\zeta>0$ and $\zeta=(1/2)\min{\mu}{\nu}$.
The fact that $\zeta\in]0,1[$ when $\mean{m^2}=\infty$ or $\mean{\K^{-2}}=\infty$, shows that the eigenmodes delocalize slower as $\omega\to0$, compared to the ``normal'' case ($\mean{m^2}<\infty$ and $\mean{\K^{-2}}<\infty$).
Our formalism has shown that the low frequency properties of the spring chain are controlled by the moments of $m_n$ and $1/\K_n$, or by the exponents of the power law tails of their distributions, when the moments are divergent.
For this reason, our results are universal.

\begin{figure}[!h]
\centering
\includegraphics[width=0.4\textwidth]{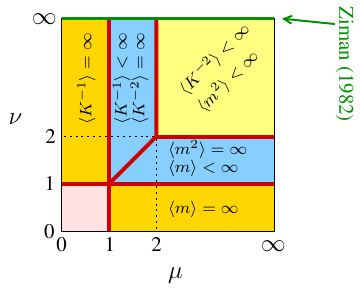}
\caption{\it Phase diagram for the low frequency spectral and localization properties of the random spring chain model in the plane $(\mu,\nu)$~;
$\mu$ is the exponent of the inverse spring constant distribution and $\nu$ the exponent of the mass distribution.
Red continuous lines correspond to first order transitions (Table~\ref{tab:Zeta} gives the exponent $\zeta$ in the different regions). 
Ziman~\cite{Zim82} found the exponent $\zeta$ for a model corresponding to $\K_n$ with power-law distribution and fixed masses (corresponding to $\nu=\infty$)~: green line.}
\label{fig:PhDiag}
\end{figure}

For power-law distributions, our method has allowed us to unveil the full phase diagram for the low frequency spectral and localization properties of random spring chains~: 
we have recovered few known results and obtained several new ones.
In particular we have identified that the spectral density exhibits two first order phase transition lines for $\mu=1$ or $\nu=1$, i.e. at the boundary of the region where $\smean{m_n}<\infty$ and $\smean{K_n^{-1}}<\infty$ (note that the Lyapunov exponent plays the role of the free energy per volume in several statistical physics' models, hence it makes sense to interpret these discontinuities as phase transitions).
The Lyapunov exponent also exhibits first order phase transitions on these lines and additional ones at the boundaries of the region where $\smean{m_n^2}<\infty$ and $\smean{K_n^{-2}}<\infty$.
The phase diagram of the Lyapunov exponent is represented in Fig.~\ref{fig:PhDiag}. 
Let us stress that the phase transitions at $\mu=2$ or $\nu=2$ can also be identified by inspection of the subleading contribution of the IDoS, see Eq.~\eqref{eq:206}, which obviously follows from the analyticity of the complex Lyapunov exponent.
The phase diagram shows the regions where normal weak disorder regime holds~:
standard weak-disorder expansions only provide the leading terms of the DoS and the Lyapunov exponent in the yellow region, when $\mu>2$ and $\nu>2$. With our new method, we were able to tackle the problem in the full plane $(\mu,\nu)$.

The study of the coefficient $a_{n+1}^{(k)}$ has relied on a detailed analysis of symmetrized products of $k$ random variables among $n$, of the form $\ssymmetriz{\prod_{i=1}^kX_{\pi(i)}}_{\pi\in\mathcal{S}_n}$, where the bracket stands for full symmetrization (i.e. uniform averaging over the permutations of the symmetric group $\mathcal{S}_n$).
We have determined the typical behaviour of such products in the limit $n\to\infty$ with $k/n$ fixed, for various types of distributions of the i.i.d. random variables $X_j$'s. For power law $p(x)=\mu\,\heaviside(x-1)\,x^{-\mu-1}$ where $\heaviside(x)$ is the Heaviside function, we obtained the typical value for $k\ll n$~:
\begin{align}
  &
  \symmetriz{\prod_{i=1}^kX_{\pi(i)}}_{\pi\in\mathcal{S}_n} 
  \\\nonumber
  &
  \sim 
  \begin{cases}
    \left( \frac{\pi\mu}{\sin\pi\mu} \right)^{k/\mu} \left( \frac{n\,\EXP{}}{k} \right)^{k(1-\mu)/\mu}
    & 
    \mbox{for } \mu \in ]0,1[ 
    \\[0.125cm]
    \left( \ln \frac{n}{k} \right)^k
    & \mbox{for } \mu=1 
    \\[0.125cm]
    \mean{X}^k \exp\left\{  -b_\mu\, \frac{k^\mu}{n^{\mu-1}} \right\}
    & 
    \mbox{for } \mu \in ]1,2[ 
    \\[0.125cm]
    \mean{X}^k \exp\left\{  -\frac{k^2}{4n}\,\ln(2n/\EXP{}k) \right\}
    & \mbox{for } \mu=2 
    \\[0.125cm]
    \mean{X}^k \exp\left\{ -\frac{\mathrm{var}(X)}{2\mean{X}^2}\,\frac{k^2}{n} \right\}
    & \mbox{for } \mu>2 
  \end{cases}
\end{align}
with $b_\mu = -\big[(\mu-1)/\mu\big]^\mu\pi/\sin(\pi\mu)>0$.
We stress that we are giving here the \textit{typical} behaviour of the symmetrized product of random variables, i.e. the average of its logarithm.
These results were obtained using the mapping onto the canonical partition function of $k$ fermions occupying $n$ random energy levels.
The fact that fluctuations of the log are negligible appears as a consequence of some standard extensivity property.
Let us explain this observation from another point of view, which will allow us to make some connection with multiple zeta functions 
$$  
  \zeta(s_1,\cdots,s_k)
  =
  \sum_{1\leq n_1<n_2<\cdots<n_k}
  n_1^{-s_1} n_2^{-s_2} \cdots n_k^{-s_k}
$$
studied in \cite{BorBraBro97}.
In Eq.~\eqref{eq:ZforOrderedIndices}, the fluctuations of the ordered $n\to\infty$ random variables $X_1<X_2<\cdots<X_n$ are reduced by the constaints.
This leads to the idea to replace the variables by the quantiles $X_i^*$ defined by $\mathcal{I}(X_i^*)=i/n$, 
where $\mathcal{I}(x)=\int^x\D u\,p(u)$ is the cumulative distribution.
Consider now the \textit{non random} product 
$\tilde Z(k,n) =  \sum_{1\leq i_1 < i_2 <\cdots < i_k\leq n}  \ \prod_{m=1}^k X_{i_m}^*$.
We can apply the same technique as in paragraph~\ref{subsec:SymmetrizedProducts}, leading to the expression of the mean occupation of levels  $\theta=k/n=(1/n)\sum_{i=1}^n\varphi X_i^*/\big[1+\varphi X_i^*\big]$, which now involves non random quantities.
In the limit $n\to\infty$, we can replace the sum by an integral, $\sum_i\to\int\D i$ and use $\D i/n=\D X_i^*\,p(X_i^*)$, leading to \eqref{eq:SaddleForTheta} and therefore showing that $Z(k,n)\approx\tilde Z(k,n)$. 
For power-law distributions considered in this paper, the quantiles are $X_i^*=(n/i)^{1/\mu}$, leading to
$\tilde Z(k,n)= n^{k/\mu}\sum_{1\leq i_1 < i_2 <\cdots < i_k\leq n}i_1^{-1/\mu}\cdots i_k^{-1/\mu}$.
When $\mu\in]0,1[$, the multiple sum converges when $n\to\infty$~: up to the factor $n^{k/\mu}$,  \eqref{eq:ZknForInfiniteMeanX} thus corresponds to the asymptotics of the multiple zeta function for degenerate indices 
$$
\zeta(\{1/\mu\}_k)
  \sim
  \left( \frac{ \pi\mu\EXP{}}{k\sin\pi\mu}\right)^{k/\mu}
\:.
$$
This behaviour coincides with the one given in \cite{BorBraBro97} (Eq.~48 of this paper).
We have therefore provided a very simple derivation of this asymptotic formula.

Finally, let us mention that the fermionic random energy model introduced in paragraph~\ref{subsec:SymmetrizedProducts} has recently attracted attention~\cite{SchHarMajSch18}, with a focus on the distribution of the ground state energy. 
In our paper, we only needed the averaged free energy of the fermions, however it would be interesting to extend the result of \cite{SchHarMajSch18} to finite temperatures and study the fluctuations of the free energy of the $k$ fermions.


In the paper, we have considered the case of i.i.d. random variables (masses and spring constants).
The case of correlated disorder has also received attention, for example in Refs.~\cite{Rus02,DeMCouRapLyr03} (see \cite{IzrKroMak12} for a broader perspective on 1D correlated disorder). 
More recently, a nice perturbative formula was obtained \cite{HerMen23} for the Lyapunov exponent, expressed in terms of mass and spring constant correlators~; the formula can however not been used for power law disorder when the second moment of the disorder is infinite.
It would interesting to make use of our exact solution (\ref{eq:MainRes0},\ref{eq:MainResIntro2}) in this case.


\section*{Acknowledgments}

The present work has emerged from the study of a model considered in the article~\cite{BerLeDRosTex24} co-authored with our colleagues Alberto Rosso and Pierre Le Doussal, who we acknowledge.


\appendix

\section{Coefficients $a_{n+1}^{(k)}$ for Dirichlet/Dirichlet boundary conditions}
\label{app:DirDir}

We now consider the situation where the chain is pinned at the two sides~:
\begin{equation}
  x_0 = x_{N+1} = 0
\end{equation}
with $\K_0\neq0$ and $\K_N\neq0$, and construct the solution $x_n(\lambda)$ of the initial value problem in this case.
In this case, it is convenient to choose the initial condition
\begin{align}
  x_1=1/\K_0
\end{align} 
Then, we have $a_1^{(0)}=1/\K_0$.
Since $a_0^{(0)}=0$, the solution of the recurrence~\eqref{eq:ReccComp0} is obviously $u_{n+1}^{(0)}=1/\K_n$, i.e. 
\begin{equation}
 \label{eq:SolutionAt0}
 x_{n+1}(0) = a_{n+1}^{(0)} = \sum_{j=0}^n\frac{1}{\K_j}
  \:.
\end{equation}
For $k\geq1$, the equation \eqref{eq:20} is solved in the same way as in the text, leading to \eqref{eq:IterationU}. 
We now get 
\begin{align}
   u_{n+1}^{(1)}  &=  \frac{1}{K_n}\sum_{i_1=0}^{n-1} \frac{\sum_{j=i_1+1}^{n} m_j }{\K_{i_1}} 
   \\
   u_{n+1}^{(2)}  &=  \frac{1}{K_n}\sum_{i_2=1}^{n-1} \sum_{i_1=0}^{i_2-1} 
    \frac{\sum_{j=i_1+1}^{i_2} m_j }{K_{i_1}}\,
    \frac{\sum_{j=i_2+1}^{n-1} m_j }{K_{i_2}}
 \end{align} 
 etc.
We deduce 
$a_{n+1}^{(k)}=\sum_{p=k}^n u_{p+1}^{(k)}$~:
\begin{equation}
  a_{n+1}^{(k)} 
  = 
  \hspace{-0.5cm}
  \sum_{0\leq i_1<\cdots<i_{k+1}\leq n}
  \frac{1}{\K_{i_{k+1}}}
  \prod_{m=1}^k \bigg(\frac{1}{\K_{i_{m}}}\sum_{j=i_m+1}^{i_{m+1}} m_j \bigg)
\end{equation}
We can also rewrite the sums more explicitly as 
\begin{equation}
  a_{n+1}^{(k)} 
  = 
  \hspace{-0.5cm}
  \sum_{0\leq i_1<j_1\leq i_2<j_2\leq\cdots \leq i_k<j_k\leq i_{k+1}\leq n}
  \frac{1}{\K_{i_{k+1}}}
  \prod_{m=1}^k \frac{m_{j_m}}{K_{i_m}}
\end{equation}
Compared to the form \eqref{eq:MainRes} obtained for Neumann/Dirichlet boundary conditions, there is one more sum and one more spring constant.
Also, the mass indices and the spring constant indices appear each time in the reverse order.

Finally, ordering differently the sums, we can get the counterpart of the representation \eqref{eq:MainRes1}
\begin{equation}
  \label{eq:CoeffForDirDir}
  a_{n+1}^{(k)} = 
  \hspace{-0.5cm}
  \sum_{1\leq j_1 <\cdots<j_{k}\leq i_{k+1}\leq n}  
  \frac{1}{\K_{i_{k+1}}}
  \prod_{m=1}^{k} \bigg( m_{j_{m}}  
  \sum_{i=j_{m-1}}^{j_{m}-1} \frac{1}{\K_i}\bigg)
\end{equation}
with $j_0=0$. 

We see that the coefficient of the $(-\lambda)^n$ term in $x_{n+1}(\lambda)$ is 
\begin{equation}
  a_{n+1}^{(n)} = \frac{1}{\K_0}\prod_{j=1}^N\frac{m_j}{\K_j} 
\end{equation}
hence the solution of the initial value problem is related to the (Dirichlet/Dirichlet) spectral determinant as 
\begin{equation}
  x_{N+1}(\lambda)  =  \frac{1}{\K_0}\prod_{j=1}^N\frac{m_j}{\K_j}
  \:\det\left( \Lambda - \lambda\,\mathbf{1}_N \right)
  \:.
\end{equation}
Compared to the Neumann/Dirichlet case (when $\K_0=0$), Eq.~\eqref{eq:SpectralDet-Neumann/Dirichlet}, there is an additional factor $1/\K_0$.

\section{Continuous models}

\subsection{Elastic model}

Writing the equation \eqref{eq:WaveEquation2} under the form 
\begin{equation}
 -m_n\lambda\, y_n 
 = \K_{n}\, (y_{n+1} - y_n)  -  \K_{n-1}\, (y_n - y_{n-1}) 
\end{equation}
makes clear what is the continuum limit of the model~:
set $x=n\,\epsilon$, where $\epsilon$ is a lattice spacing, $y(x)=y_n$, $k(x)=\epsilon^2\K_n$ and $m(x)=m_n$.
Assuming that $k(x)$ and $m(x)$ are two smooth functions at the scale of $\epsilon$, in the limit $\epsilon\to0$ we get
\begin{equation}
  \label{eq:ElasticModel}
  - m(x)\lambda\, y(x) = \partial_x  k(x)\partial_x  y(x) 
  \:.
\end{equation}
We adapt the derivation of Section~\ref{sec:Combinatorics} to this continuous model.
We search the solution under the form
\begin{equation}
  y(x) = \sum_{k=0}^\infty (-\lambda)^k \, a_{k}(x)
\end{equation}
for initial conditions $y(0)=0$ and $y'(0)=1/k(0)$.
The coefficient $a_0(x)$ solves the homogeneous equation, leading to
\begin{equation}
  a_0(x) = \int_0^x\frac{\D u}{k(u)}
\end{equation}
and fulfills the two initial conditions $y(0)=a_0(0)=0$ and $y'(0)=a_0'(0)=1/k(0)$. The other coefficients solve the recurrence
\begin{equation}
  \partial_x k(x)\partial_x  a_{k}(x) = m(x)\,a_{k-1}(x)
\end{equation}
for initial condition $a_{k}(0)=a_{k}'(0)=0$.
We find 
\begin{equation}
  a_{k}(x) = \int_0^x\D t \,a_{k-1}(t)\, m(t) \int_t^x\frac{\D u}{k(u)}
  \:,
\end{equation}
hence we can write
\begin{align}
  \label{eq:CoeffElasticModel}
  a_{k}(x) &= 
  \hspace{-0.5cm}
  \int\limits_{0<t_1<\cdots<t_k<x}
  \hspace{-0.5cm}
  \D t_1\cdots\D t_k
  \prod_{p=0}^k  m(t_p)\int_{t_p}^{t_{p+1}}\frac{\D u_p}{k(u_p)} 
  \\
  & = 
 \hspace{-1.25cm}
  \int\limits_{0<u_0<t_1<u_1<\cdots<t_k<u_k<x}
  \hspace{-1.25cm}
  \D u_0\D t_1\D u_1\cdots\D t_k\D u_k
  \prod_{p=0}^k  \frac{m(t_p)}{k(u_p)} 
\end{align}
where $t_0=0$, $t_{k+1}=x$ and $m(0)=1$ have made the formula more compact.
This is the continuous version of Eq.~\eqref{eq:CoeffForDirDir}.

\subsection{Supersymmetric model}

Interesting features of the Anderson model with random hoppings \eqref{eq:AndersonRandomCouplings} occur in the band center ($E=0$).
For hoppings of the form $\rh_n=[1/\epsilon -(-1)^nm(x)]/2$ where $x=(n+/2)\epsilon$, $\epsilon\to0$ being a lattice spacing, the Anderson model for small energy is mapped onto the Dirac equation for $\widetilde{\mathcal{H}}_D=-\I\sigma_z\partial_x+\sigma_y\,m(x)$.
We prefer to consider the rotated Hamiltonian $\mathcal{H}_D=\I\sigma_y\partial_x+\sigma_x\,m(x)$, whose square involves a pair of supersymmetric partner Hamiltonians $H_+=Q^\dagger Q$ and $H_-=QQ^\dagger$, with $Q=-\partial_x+m(x)=-\EXP{-U(x)} \partial_x\EXP{U(x)}$.
Localization for the supersymmetric Hamiltonian 
\begin{equation}
  \label{eq:Hsusy}
  H_+ = Q^\dagger Q = -\EXP{U(x)}\partial_x\EXP{-2U(x)}\partial_x\EXP{U(x)} 
  \:,
\end{equation}
where $U(x)=-\int_0^x\D y\,m(y)$,
has been much studied \cite{BouComGeoLeD90,Tex00,ComTexTou11,ComTexTou13} (for reviews, cf. Refs.~\cite{ComTex98,TexHag10}).
When $\mean{m(x)}=0$, the model is well-known to present the Dyson singularity $\rho(E)\sim1/(E|\ln E|^3)$ recovered in Section~\ref{sec:DysonTypeI} for the discrete Anderson model.
We apply the method for the construction of the initial value problem
\begin{equation}
  H_+ \psi(x;E) = E \,\psi(x;E)
\end{equation}
for initial conditions $\psi(0;E)=0$ and $\psi'(0;E)=1$.
We expand the solution in powers of the spectral parameter
\begin{equation}
  \psi(x;E) = \sum_{k=0}^\infty (-E)^k \, a_{k}(x)
\end{equation}
where the coefficients obey the recurrence 
\begin{equation}
  \label{eq:Susy-Recu-ak}
  H_+ a_{k}(x) = - a_{k-1}(x)
\end{equation}
with $a_{k}(0)=0$ and $a_{k}'(0)=\delta_{k,0}$.
Using the specific structure of the Hamiltonian $H_+$, we easily obtain the two independent solutions of $H_+\psi=0$~:
\begin{equation}
  \psi_0(x) = \EXP{-U(x)} 
  \hspace{0.25cm}\mbox{and}\hspace{0.25cm}
  \psi_1(x) = \psi_0(x)\int_0^x\frac{\D y}{\psi_0(y)^2}
  \:.
\end{equation}
Clearly $Q\psi_0(x)=0$ and $Q\psi_1(x)=-1/\psi_0(x)$.
Making use of these remarks, we get $a_0(x) = \psi_1(x)$ and
\begin{equation}
  a_k(x) = \psi_1(x)\int_0^x\D y\,\psi_0(y)\,a_{k-1}(y)
  \:.
\end{equation}
Iteration gives 
\begin{equation}
  a_k(x) = \psi_1(x)
  \hspace{-0.5cm}
  \int\limits_{0<y_1<\cdots<y_k<x}
  \hspace{-0.5cm}
  \D y_1\cdots\D y_k\,
\prod_{p=1}^k \psi_0(y_p)\,\psi_1(y_p)
\end{equation}
with $\psi_0(y)\,\psi_1(y)=\EXP{-2U(x)}\int_0^y\D t\, \EXP{2U(t)}$. 
The structure of the coefficients is similar to the one of the elastic model, Eq.~\eqref{eq:CoeffElasticModel}~:
this can be understood by comparing \eqref{eq:Hsusy} with the operator 
\begin{equation}
\Lambda=-m(x)^{-1}\partial_x k(x)\partial_x
\end{equation}
 involved in the elastic model~\eqref{eq:ElasticModel}. Clearly, there is a correspondence when $m(x)=k(x)=\EXP{-2U(x)}$.
 
 \medskip
 
\paragraph*{Remark~:} 

the calculus of the coefficients $a_k(x)$ share some similarity with the problem of first passage time of a diffusive particle submitted to a drift $F(x)=-U'(x)$.
Consider such a particle starting from $x_0$.
We study the time of first passage at $x=b>x_0$. 
It is well-known that the $k$-th moment of the first passage time, denoted $T_k(x_0)$, obeys the backward Fokker-Planck equation with a source term given by the $(k-1)$-th moment
\begin{equation}
\label{eq:Recu-FirstPassage}
  \mathscr{G}_x T_k(x) = -k\,T_{k-1}(x)
\end{equation}
with boundary condition $T_k(b)=0$. 
The generator of the diffusion can be written as $\mathscr{G}_x=(1/2)\partial_x^2-U'(x)\partial_x=(1/2)\EXP{2U(x)}\partial_x\EXP{-2U(x)}\partial_x$, where  the diffusion constant is $D=1/2$.
Furthermore, choosing a reflecting boundary condition $T_k'(0)=0$ at $x=0<x_0$,  we get
\begin{equation}
  T_k(x_0) = 2k \int_{x_0}^b \D x \,\EXP{2U(x)} \int_0^x\D y\,\EXP{-2U(y)} \, T_{k-1}(y)
\end{equation}
see the monograph \cite{Gar89} (or Appendix of \cite{Tex00} for a short reminder).
These moments were studied for a disordered potential in \cite{CasLeD01}.

Interestingly, the recurrence \eqref{eq:Susy-Recu-ak} is related to the recurrence \eqref{eq:Recu-FirstPassage}.
Introducing $\tau_k(x) = (-2)^k\,k!\,\EXP{U(x)}\,a_k(x)$ we indeed get 
$\mathscr{G}_x \tau_k(x) = -k\,\tau_{k-1}(x)$.
However the difference between the two problems lies in the boundary conditions~:
the conditions given above for the coefficients $a_k(x)$'s rewrite
$\tau_k(0)=0$ and $\partial_x\big[\EXP{-U(x)}\tau_k(x)\big]\Big|_{x=0}=\delta_{k,0}$, hence there is no correspondence with the moments of the first passage time (in particular $\tau_0(x)=\EXP{U(x)}a_0(x)=\int_0^x\D y\,\EXP{2U(y)}\neq1$).


\section{Few sums}
\label{app:Sums}

In this appendix, we compute several sums which appeared above.

\subsection{The sum $S_k(n)$ (for the case $\mu>1$ and $\nu>1$)}

We study the sum \eqref{eq:DefSnk}. 
We introduce the variables $q_m=i_{m}-i_{m-1}\geq1$. 
The constraint $i_{k}\leq n$ is now $q_1+\cdots+q_k\leq n$, hence we can write the sum as
\begin{equation}
  S_k(n) = \sum_{m=k}^n \sigma_k(m)
\end{equation}
with 
\begin{equation}
  \sigma_k(n) = \sum_{q_1=1}^\infty\cdots\sum_{q_k=1}^\infty  \delta_{n,q_1+\cdots+q_k} \prod_{m=1}^{k} q_m
  \:.
\end{equation}
The constraint can be removed by considering the generating function 
\begin{align}
  Z_k(s) &= \sum_{n=0}^\infty s^n \,\sigma_k(n)
  =  \bigg( \sum_{q=1}^\infty q\,s^q \bigg)^k 
  = \frac{s^k}{(1-s)^{2k}}
  \:.
\end{align}
The sum $\sigma_k(n)$ can be represented as a contour integral in the complex plane, 
$\sigma_k(m)=(2\I\pi)^{-1}\oint\D s\,s^{-m-1}\,Z_k(s)$, with contour encircling once the origin.
Sum over~$m$ gives
\begin{align}
\nonumber  
S_k(n) &= \sum_{m=k}^n \oint\frac{\D s}{2\I\pi} \frac{Z_k(s)}{s^{m+1}}
=  \oint\frac{\D s}{2\I\pi}\,Z_k(s) \frac{s^{-n-1}-s^{-k}}{1-s}
  \\
  &=\oint\frac{\D s}{2\I\pi}\frac{1}{(1-s)^{2k+1}} \left[\frac{1}{s^{n-k+1}} - 1 \right]
\end{align}
Because the second term is analytic at $s=0$, its contribution to the integral vanishes.
Using  $(1-s)^{-\alpha}=\sum_{n=0}^\infty\big[(\alpha)_n/n!\big]\,s^n$, 
where $(\alpha)_n=\Gamma(\alpha+n)/\Gamma(\alpha)$ is the Pochhammer symbol, we finally obtain
\begin{equation}
  \label{eq:SumSnk}
  S_k(n) = \frac{(n+k)!}{(n-k)!(2k)!}
  =\begin{pmatrix}
    n+k \\ 2k
  \end{pmatrix}
\end{equation}
Asymptotics for $n\gg k\gg1$ is easily found from the Stirling formula
\begin{equation}
  \label{eq:SumSnkAsymp}
  S_k(n)   \simeq \frac{1}{\sqrt{4\pi k}} \left(\frac{n\,\EXP{}}{2k}\right)^{2k}
  \:.
\end{equation}

\subsection{General method}
\label{app:ComputationOfNiceSums}

We generalize the method of the previous subsection in order to deal with other sums involved in the paper.
We consider a sum of the form 
\begin{equation}
  \Sigma_{k}(n) =\sum_{1\leq i_1 < i_2 < \cdots < i_{k} \leq n}  \prod_{m=1}^{k} f(i_{m}-i_{m-1}) 
\end{equation}
with $i_0=0$, where $f(x)$ is some function.

(i)
In a first step, we rewrite the sum under the form
\begin{equation}
  \Sigma_{k}(n) = \sum_{m=k}^n \sigma_{k}(m)
\end{equation}
where 
\begin{equation}
   \sigma_{k}(n) =  \sum_{q_1=1}^\infty\cdots\sum_{q_k=1}^\infty  \delta_{n,q_1+\cdots+q_k} \prod_{m=1}^{k} f(q_m)
\end{equation}
Note that this expression shows that $\sigma_{k}(n)=0$ for $n<k$.

(ii)
In a second step we introduce the generating function 
\begin{equation}
  Z_{k}(s) = \sum_n s^n \sigma_{k}(n) =  \big[ \check{f}(s) \big]^k  
\end{equation}
where
\begin{equation}
  \check{f}(s) = \sum_{j=1}^\infty f(j) \, s^j
\end{equation}
The function $f$ should be such that the series is convergent at least in a small domain around $s=0$.
In the continuum, $\check{f}$ would be the Mellin transform of~$f$.

(iii)
We represent $\sigma_{k}(m)$ in term of the contour integral and sum over $m$ to get $\Sigma_{k}(n) $~:
\begin{align}
  \Sigma_{k}(n)  = \sum_{m=k}^n \oint\frac{\D s}{2\I\pi} \frac{Z_{k}(s)}{s^{m+1}}
  = \oint\frac{\D s}{2\I\pi} \frac{\big[ \check{f}(s) \big]^k(s^{-n-1}-s^{-k})}{1-s}
\end{align}
where the contour of integration encircles once the origin, in the complex plane.
Since $\check{f}(s)\sim s$ for $s\to0$, the function $[ \check{f}(s)]^k s^{-k}/(1-s)$ is analytic at $s=0$, hence do not contribute to the contour integral, thus  
\begin{align}
  \Sigma_{k}(n) = \oint\frac{\D s}{2\I\pi}\, \frac{\big[ \check{f}(s) \big]^k}{s^{n+1}(1-s)}
  \:.
\end{align}

(iv) The last step~: asymptotic of the sum for large $n$ is obtained thanks to some saddle point approximation~:
\begin{equation}
  \label{eq:SigmaknAsymp}
   \Sigma_{k}(n) \sim \frac{\big[ \check{f}(s_*) \big]^k}{s_*^{n+1}(1-s_*)}
   \hspace{0.25cm}\mbox{for } n\to\infty
   \:,
\end{equation} 
 where $s_*$ is the solution of 
\begin{equation}
  \label{eq:SaddlePoint}
  \deriv{}{s}
  \big[
    (n+1)\ln(1/s) - \ln(1-s) + k\,\ln \check{f}(s)
  \big]   =0
  \:.
\end{equation} 
In the limit $n\gg k,\:1$, we expect the sum to be dominated by a saddle point $s_*$ close to $1$.

\subsubsection{First illustration}

We compute the sum
\begin{equation}
  \label{eq:DefRnk}
  R_{k,\mu}(n) = 
  \hspace{-0.5cm}  
  \sum_{1\leq i_1<i_2 <\cdots<i_{k}\leq n} \prod_{m=1}^{k} (i_{m}-i_{m-1})^{1/\mu}
  \:.
\end{equation}
We follow the method of the paragraph~\ref{app:ComputationOfNiceSums}.
In this case the function is $f(x)=x^{1/\mu}$, hence
\begin{equation}
  \check{f}(s) = \sum_{j=1}^\infty j ^ {1/\mu} \, s^j
  = \mathrm{Li}_{1/\mu}(s)
\end{equation}
is the polylogarithm function \cite{DLMF}. Because we need $\check{f}(s)$ in the $s\to1^-$ limit, we can more simply write
\begin{equation}
  \label{eq:intB13}
  \check{f}(s)\simeq \int_0^\infty \D t\, t^ {1/\mu} s^t
  = \frac{\Gamma(1/\eta)}{[\ln(1/s)]^{1/\eta}}
\end{equation}
with $\eta=\mu/(1+\mu)$.
The saddle point equation~\eqref{eq:SaddlePoint} takes the explicit form 
\begin{equation}
  n+1  - \frac{s_*}{1-s_*} - \frac{k}{\eta}\frac{1}{\ln(1/s_*)} \simeq0
  \:,
\end{equation}
hence 
\begin{equation}
  s_* \simeq 1 -\frac{1}{n} \left( 1 + k/\eta \right)
  \hspace{0.5cm}\mbox{for }
  n\gg1
  \:.
\end{equation}
(the validity of the assumption $s_*\to1^-$ also requires $n\gg k$).
Finally, from \eqref{eq:SigmaknAsymp}, we get 
\begin{equation}
  \label{eq:RknAsymp}
  R_{k,\mu}(n) \sim
  \Gamma(1/\eta)^k \left(  \frac{ n \, \EXP{}}{1+k/\eta}  \right)^{k/\eta}
\end{equation}
for $\eta=\mu/(1+\mu)$.
As a check, we can consider the case $\mu=1$ ($\eta=1/2$), when $R_{k,1}(n)=S_k(n)$.
We check that the asymptotic $R_{k,1}(n)\sim[n\EXP{}/(2k)]^{2k}$ coincides with the one deduced in the first paragraph from the exact form of $S_k(n)$, Eq.~\eqref{eq:SumSnkAsymp}.

\subsubsection{Second illustration}

We now analyze the sum
\begin{equation}
  \label{eq:DefStilde-nk}
  \widetilde{S}_k(n) = 
  \hspace{-0.5cm}  
  \sum_{1\leq i_1 <\cdots<i_k \leq n} \prod_{m=1}^{k} (i_{m}-i_{m-1})\,\ln(i_{m}-i_{m-1})
  \:.
\end{equation}
We apply the method of Subsection~\ref{app:ComputationOfNiceSums} for $f(x)=x\,\ln x$, hence we consider (for $s\to1^-$)
\begin{equation}
 \check{f}(s) = \sum_{j=1}^\infty j\,s^j\,\ln j
  \simeq 
  \int_0^\infty\D t\,\ln t\,s^t
  =\frac{1-\mathbf{C}-\ln\ln(1/s)}{[\ln(1/s)]^2}
\end{equation}
where $\mathbf{C}\simeq0.577$ is the Euler-Mascheroni constant (we can obtain this integral from \eqref{eq:intB13} by setting $1/\eta=2+\epsilon$, eventually identifying the $\mathcal{O}(\epsilon)$ term).
It is convenient to introduce $b=\EXP{1-\mathbf{C}}\simeq1.526$.
The saddle point equation~\eqref{eq:SaddlePoint} takes the form 
\begin{equation}
  n+1  - \frac{s_*}{1-s_*} - \frac{k}{\ln(1/s_*)\ln(b/\ln(1/s_*))} -\frac{2k}{\ln(1/s_*)} \simeq0
\end{equation}
leading to 
\begin{equation}
  s_* \simeq 1 - \frac{1}{n}\left( 2k+1 + \frac{k}{\ln(bn/2k)} \right)
\end{equation}
for $n\gg1 $.
Finally we obtain 
\begin{align}
  \widetilde{S}_k(n) \sim
  &
  \left(\frac{n}{ 2k+1+\frac{k}{\ln(bn/2k)} }\right)^{2k+1}
  \left(\ln \frac{b\,n}{2k}\right)^k
  \nonumber\\
  &\times\exp\left\{ 2k+1+\frac{k}{\ln(bn/2k)} \right\}
\end{align}
For $n\gg k\gg1$, it simplifies as 
\begin{equation}
  \label{eq:AsympStilde}
  \widetilde{S}_k(n) \sim \left(\frac{n\,\EXP{}}{2k}\right)^{2k} \left(\ln \frac{b\,n}{2k}\right)^k
  \:.
\end{equation}


\section{The $\lambda\to-\infty$ limit for exponential distributions}
\label{app:ExpCase-ExpansionOmega-Large}

As we have noticed, for exponential distribution of $\K_n^{-1}$ and $m_n$, the two fugacities coincide. 
Then \eqref{eq:MainResultForOmegaBis} can be more conveniently written
\begin{align}
\label{eq:OmegaAppendix}
  &\Omega = 
  \underset{\theta}{\mathrm{max}} \bigg\{ 
  \ln\left( \Lambda\,\EXP{-2\mathbf{C}}/4\right)
  +
   (1+\theta)\ln(1+\theta)
  \\\nonumber
  &+ (1-\theta) \left[\ln(1-\theta) - \ln(\Lambda/4)\right]
  + 2 \int_{\theta}^1 \D t\, \ln \psi(t)
  \bigg\}
\end{align}
where $\Lambda=(-\lambda)/(ab)$.
The fugacity is given by solving the transcendental equation \eqref{eq:ExpCase-EqForPsi}. 

In a first step, we study the fugacity for $\theta\to1^-$, i.e. $\psi\to\infty$, leading to 
\begin{equation}
  \psi \simeq \frac{ \ln(\psi\,\EXP{-\mathbf{C}}) }{\psi} 
  \left[ 1 + \frac{1}{\psi} + \mathcal{O}\left(\frac{1}{\psi\ln\psi}\right) \right]
\end{equation}
leading to the expansion
\begin{align}
  \psi(\theta) \simeq \frac{\ln(\EXP{-\mathbf{C}}/\varepsilon)}{\varepsilon}
  \left[ 
    1 + \frac{\ln \ln(\EXP{-\mathbf{C}}/\varepsilon)}{\ln(\EXP{-\mathbf{C}}/\varepsilon) } + \mathcal{O}\left(\frac{\ln\ln}{\ln^2}\right)
  \right]
\end{align}
where $\varepsilon=1-\theta\to0^+$. 
A careful integration in \eqref{eq:OmegaAppendix} gives
\begin{align}
\label{eq:ProvisoryOmegaAppendix}
\Omega \simeq 
&\ln\left( \Lambda\,\EXP{-2\mathbf{C}}\right)
+ 
\underset{\varepsilon}{\mathrm{max}} \bigg\{ 
\varepsilon\,
\bigg[
1 + \ln\left( \frac{2}{\Lambda\,\varepsilon}\ln^2(\EXP{-\mathbf{C}}/\varepsilon) \right)
\nonumber\\
&\hspace{1cm}
 + 2 \frac{1+\ln \ln(\EXP{-\mathbf{C}}/\varepsilon)}{\ln(\EXP{-\mathbf{C}}/\varepsilon) } 
  + \mathcal{O}\left(\frac{\ln\ln}{\ln^2}\right)
\bigg]
\bigg\}
\end{align}
The position of the maximum ($\varepsilon=1-\theta_*(\lambda)$) is solution of \eqref{eq:ExpCase-SaddleEq} i.e.
\begin{align}
 \label{eq:D5}
  \varepsilon 
  \simeq
   \frac{2\ln^2\left( \EXP{-\mathbf{C}} / \varepsilon\right)}{\Lambda}\,
  \left[
     1 +2 \frac{\ln\ln\left( \EXP{-\mathbf{C}} / \varepsilon\right)}{\ln\left( \EXP{-\mathbf{C}} / \varepsilon\right) }
     + \mathcal{O}\left( \frac{(\ln\ln)^2}{\ln^2}\right)
  \right] 
\end{align}
Replacing $\varepsilon$ by its expression in the logarithms of the r.h.s., we get 
\begin{align}
  \varepsilon \simeq \frac{2}{\Lambda}
  \bigg[
    \ln^2(\star) - 2 \ln(\star)  \, \ln\ln (\star) + \mathcal{O}\left((\ln\ln)^2\right)
  \bigg]
\end{align}
where 
$(\star)=(\Lambda\EXP{-\mathbf{C}}/2)$.
Making use of \eqref{eq:D5}, we can simplify \eqref{eq:ProvisoryOmegaAppendix} as 
\begin{equation}
  \label{eq:LastEqForOmega-exp}
  \Omega \simeq 
\ln\left( \Lambda\,\EXP{-2\mathbf{C}}\right)
+ 
 \varepsilon\,\left\{
 1
 + \frac{2}{\ln\left(\EXP{-\mathbf{C}} /  \varepsilon \right) }
     + \mathcal{O}\left( \frac{\ln\ln}{\ln^2}\right)
 \right\}
\end{equation}
where it is now understood that $\varepsilon$ solves \eqref{eq:D5}. Using the above expression of $\varepsilon$, we get  \eqref{eq:OmegaForExpWeights}.



\begin{thebibliography}{10}
\bibitem{footnote2} 
  In his pioneering work, Dyson's starting point was the same spectral problem for random masses, however he solved the model in a case where masses and spring constants are correlated (called "type I" disorder, in his paper), corresponding to the tight-bingding model with random hoppings (Section~\ref{sec:DysonTypeI}). 
  For a in-depth discussion of Dyson's pioneer paper, see the recent review~\cite{For21}.
\bibitem{footnote3}
  In Ref.~\cite{Nie84} (\S~4), Nieuwenhuizen identified a solvable case where 
    $1/\K_n$ has exponential distribution $p(x)=a\,\EXP{-ax}$ and $m_n$ has distribution $q(m)=(1-p-q)\,\delta(m)+p\,b\,\EXP{-bm}+q\,b^2\,m\,\EXP{-bm}$ where $p\:,q,\:p+q\in[0,1]$.
\bibitem{footnote4}
  Other combinatorial methods have been developed in the context of disordered systems to analyze Lifshitz tails in Refs.~\cite{KucSad98,BapSem11}.
\end{thebibliography}
%

\end{document}